\documentclass[conference,compsoc]{IEEEtran}
\AtBeginDocument{%
  }
    
\usepackage[utf8]{inputenc}
\usepackage{amsmath}
\usepackage{pifont}
\usepackage{wasysym}
\usepackage{makecell}
\usepackage{multirow}
\usepackage[figuresleft]{rotating}
\usepackage{longtable,supertabular,booktabs,tablefootnote}
\usepackage{threeparttablex}
\usepackage{upgreek}
\usepackage{ltablex}
\usepackage[nolist]{acronym}

\usepackage{amsthm}
\usepackage[normalem]{ulem}
\usepackage{bm}
\usepackage{enumitem}
\usepackage{algorithm}
\usepackage[noend]{algpseudocode}
\usepackage{adjustbox}
\usepackage{colortbl}
\usepackage{graphicx}
\usepackage{soul}

\usepackage{tikz}
\usetikzlibrary{positioning,fit,backgrounds}
\definecolor{victimred}{HTML}{B33A3A}
\definecolor{attackerblue}{HTML}{2F6B9A}
\definecolor{neutralfill}{HTML}{F2F3F5}
\definecolor{victimfill}{HTML}{FBEAEC}
\definecolor{attackerfill}{HTML}{EAF2FB}
\tikzset{
  kvblock/.style={draw=black!55,line width=0.4pt,minimum height=0.52cm,
    align=center,inner sep=1.5pt,outer sep=0pt},
  kvneutral/.style={kvblock,fill=neutralfill},
  kvvictim/.style={kvblock,fill=victimfill,draw=victimred!75!black},
  kvattacker/.style={kvblock,fill=attackerfill,draw=attackerblue!75!black},
  kvrole/.style={font=\scriptsize,anchor=east},
  kvredirect/.style={->,>=stealth,line width=0.8pt,draw=victimred},
  kvfault/.style={font=\scriptsize,text=victimred,fill=white,
    inner sep=1pt,align=left}
}
\newcommand*\circled[1]{\tikz[baseline=(char.base)]{
            \node[shape=circle,draw,inner sep=0.4pt, font=\footnotesize] (char) {#1};}}

\usepackage[most]{tcolorbox}
\newtcolorbox{attackexample}{
  colback=neutralfill,colframe=black!50,boxrule=0.4pt,arc=1.5pt,
  left=4pt,right=4pt,top=3pt,bottom=3pt,
  before skip=1pt,after skip=1pt,
  segmentation style={solid,draw=black!25,line width=0.3pt},
  fontupper=\footnotesize,fontlower=\footnotesize
}
\newcommand{\victimtext}[1]{\textcolor{victimred}{\textbf{#1}}}
\newcommand{\attackertext}[1]{\textcolor{attackerblue}{\textbf{#1}}}

\usepackage[]{hyperref} 
\usepackage{cleveref}

\ifCLASSOPTIONcompsoc
  \usepackage[nocompress]{cite}
\else
  \usepackage{cite}
\fi

\ifCLASSOPTIONcompsoc
  \usepackage[caption=false,font=footnotesize,labelfont=sf,textfont=sf]{subfig}
\else
  \usepackage[caption=false,font=footnotesize]{subfig}
\fi

\usepackage{xspace}
\newcommand{\etal}{et al.\xspace}

\newcommand{\ie}{i.e.,\xspace}
\newcommand{\eg}{e.g.,\xspace}

\let\oldnl\nl%
\newcommand{\nonl}{\renewcommand{\nl}{\let\nl\oldnl}}%

\newcommand{\sys}{\textsl{\mbox{BlockJack}}\xspace}

\usepackage[textsize=scriptsize]{todonotes}
\let\oldtodo\todo
\renewcommand{\todo}[1]{\oldtodo[inline]{#1}}

\newcommand{\idx}{\mathrm{Idx}}
\newcommand{\siphon}{\ensuremath{\mathtt{CD}}\xspace}
\newcommand{\graft}{\ensuremath{\mathtt{RA}}\xspace}

\begin{document}

\title{\sys: Cross-Tenant Redirection Fault Attacks on Paged KV Cache Address Translation in LLM Serving }

\author{
\IEEEauthorblockN{
Qifan Wang\IEEEauthorrefmark{1},
Jeremy Boy\IEEEauthorrefmark{2},
Jonas Sander\IEEEauthorrefmark{2},
Yansong Gao\IEEEauthorrefmark{3},
Hongsheng Hu\IEEEauthorrefmark{4},
Ye Dong\IEEEauthorrefmark{5},\\
Thomas Eisenbarth\IEEEauthorrefmark{2},
and David Oswald\IEEEauthorrefmark{1}
}

\IEEEauthorblockA{
\IEEEauthorrefmark{1}Durham University
\quad
\IEEEauthorrefmark{2}University of L\"ubeck
\quad
\IEEEauthorrefmark{5}National University of Singapore\\
\IEEEauthorrefmark{4}Shanghai Jiao Tong University
\quad
\IEEEauthorrefmark{3}Southeast University\\[2pt]
Email: \{qifan.wang2,david.f.oswald\}@durham.ac.uk\\
\{je.boy,j.sander,thomas.eisenbarth\}@uni-luebeck.de\\
dongye@nus.edu.sg,
gao.yansong@hotmail.com,
hongsheng.hu@sjtu.edu.cn
}
}

\begin{acronym}[ASCII]
 \setlength{\itemsep}{0.2em}
 \acro{3DES}{Triple DES}
 \acro{AES}{Advanced Encryption Standard}
 \acro{ANF}{Algebraic Normal Form}
 \acro{API}{Application Programming Interface}
 \acro{ARX}{Addition, Rotate, XOR}
 \acro{ASK}{Amplitude-Shift Keying}
 \acro{ASIC}{Application Specific Integrated Circuit}
 \acro{BL}{Bootloader Enable}
 \acro{BOR}{Brown-Out Reset}
 \acro{BPSK}{Binary Phase Shift Keying}
 \acro{CBC}{Cipher Block Chaining}
 \acro{CBS}{Critical Bootloader Section}
 \acro{CGM}{Continuous Glucose Monitoring System}
 \acro{CMOS}{Complementary Metal Oxide Semiconductor}
 \acro{COPACOBANA}{Cost-Optimized Parallel Code Breaker and Analyzer}
 \acro{CPA}{Correlation Power Analysis}
 \acroplural{CPA}[CPAs]{Correlation Power Analyzes}
 \acro{CPU}{Central Processing Unit}
 \acro{CTR}{Counter \acroextra{(mode of operation)}}
 \acro{CRC}{Cyclic Redundancy Check}
 \acro{CRP}{Code Readout Protection}
 \acro{DES}{Data Encryption Standard}
 \acro{DC}{Direct Current}
 \acro{DDS}{Digital Direct Synthesis}
 \acro{DFT}{Discrete Fourier Transform}
 \acro{DMA}{Direct Memory Access}
 \acro{DoS}{Denial-of-Service}
 \acro{DFA}{Differential Fault Analysis}
 \acro{DPA}{Differential Power Analysis}
 \acro{DRAM}{Dynamic Random-Access Memory}
 \acro{DRM}{Digital Rights Management}
 \acro{DSO}{Digital Storage Oscilloscope}
 \acro{DSP}{Digital Signal Processing}
 \acro{DST}{Digital Signature Transponder}
 \acro{DUT}{Device Under Test}
 \acroplural{DUT}[DUTs]{Devices Under Test}
 \acro{ECB}{Electronic Code Book}
 \acro{ECC}{Elliptic Curve Cryptography}
 \acro{ECU}{Electronic Control Unit}
 \acro{EDE}{Encrypt-Decrypt-Encrypt \acroextra{(mode of operation)}}
 \acro{EEPROM}{Electrically Erasable Programmable Read-Only Memory}
 \acro{EM}{Electro-Magnetic}
 \acro{FFT}{Fast Fourier Transform}
 \acro{FCC}{Federal Communications Commission}
 \acro{FIR}{Finite Impulse Response} 
 \acro{FIVR}{Fully Integrated Voltage Regulator}
 \acro{FPGA}{Field Programmable Gate Array}
 \acro{FSK}{Frequency Shift Keying}
 \acro{GMSK}{Gaussian Minimum Shift Keying}
 \acro{GPIO}{General Purpose I/O}
 \acro{GPU}{Graphics Processing Unit}
 \acro{HD}{Hamming Distance}
 \acro{HDL}{Hardware Description Language}
 \acro{HF}{High Frequency}
 \acro{HMAC}{Hash-based Message Authentication Code}
 \acro{HW}{Hamming Weight}
 \acro{IC}{Integrated Circuit}
 \acro{ID}{Identifier}
 \acro{ISM}{Industrial, Scientific, and Medical \acroextra{(frequencies)}}
 \acro{IIR}{Infinite Impulse Response}
 \acro{IP}{Intellectual Property}
 \acro{IoT}{Internet of Things}
 \acro{IV}{Initialization Vector}
 \acro{JTAG}{Joint Test Action Group}
 \acro{LF}{Low Frequency}
 \acro{LFSR}{Linear Feedback Shift Register}
 \acro{LQI}{Link Quality Indicator}
 \acro{LSB}{Least Significant Bit}
 \acro{LSByte}{Least Significant Byte}
 \acro{LUT}{Look-Up Table}
 \acro{MAC}{Message Authentication Code}
 \acro{MF}{Medium Frequency}
 \acro{MMU}{Memory Management Unit}
 \acro{MITM}{Man-In-The-Middle}
 \acro{MSB}{Most Significant Bit}
 \acro{MSByte}{Most Significant Byte}
 \acro{MSK}{Minimum Shift Keying}
 \acro{MSR}{Model Specific Register}
 \acro{muC}[$\mathrm{\upmu C}$]{Microcontroller}
 \acro{NLFSR}{Non-Linear Feedback Shift Register}
 \acro{NLF}{Non-Linear Function}
 \acro{NFC}{Near Field Communication}
 \acro{NRZ}{Non-Return-to-Zero \acroextra{(encoding)}}
 \acro{NVM}{Non-Volatile Memory}
 \acro{OOK}{On-Off-Keying}
 \acro{OP}{Operational Amplifier}
 \acro{OTP}{One-Time Password}
 \acro{PC}{Personal Computer}
 \acro{PCB}{Printed Circuit Board}
 \acro{PhD}{Patiently hoping for a Degree}
 \acro{PKE}{Passive Keyless Entry}
 \acro{PKES}{Passive Keyless Entry and Start}
 \acro{PKI}{Public Key Infrastructure}
 \acro{PMBus}{Power Management Bus}
 \acro{PoC}{Proof-of-Concept}
 \acro{POR}{Power-On Reset}
 \acro{PPC}{Pulse Pause Coding}
 \acro{PRNG}{Pseudo-Random Number Generator}
 \acro{PSK}{Phase Shift Keying}
 \acro{PWM}{Pulse Width Modulation}
 \acro{RAPL}{Running Average Power Limit}
 \acro{RDP}{Read-out Protection}
 \acro{RF}{Radio Frequency}
 \acro{RFID}{Radio Frequency IDentification}
 \acro{RKE}{Remote Keyless Entry}
 \acro{RNG}{Random Number Generator}
 \acro{ROM}{Read Only Memory}
 \acro{ROP}{Return-Oriented Programming}
 \acro{RSA}{Rivest Shamir and Adleman}
 \acro{SDR}{Software-Defined Radio}
 \acro{SGX}{Software Guard Extensions}
 \acro{SNR}{Signal to Noise Ratio}
 \acro{SHA}{Secure Hash Algorithm}
 \acro{SHA-1}{Secure Hash Algorithm 1}
 \acro{SHA-256}{Secure Hash Algorithm 2 (256-bit version)}
 \acro{SMA}{SubMiniature version A \acroextra{(connector)}}
 \acro{SMBus}{System Management Bus}
 \acro{I2C}{Inter-Integrated Circuit}
 \acro{SPA}{Simple Power Analysis}
 \acro{SPI}{Serial Peripheral Interface}
 \acro{SPOF}{Single Point of Failure}
 \acro{SoC}{System on Chip}
 \acro{SVID}{Serial Voltage Identification}
 \acro{SWD}{Serial Wire Debug}
 \acro{TCB}{Trusted Computing Base} 
 \acro{TEE}{Trusted Execution Environment}
 \acro{TMTO}{Time-Memory Tradeoff}
 \acro{TMDTO}{Time-Memory-Data Tradeoff}
 \acro{TZ}[TrustZone]{TrustZone}
 \acro{RSSI}{Received Signal Strength Indicator}
 \acro{SHF}{Superhigh Frequency}
 \acro{UART}{Universal Asynchronous Receiver Transmitter}
 \acro{UCODE}[$\mathrm{\upmu Code}$]{Microcode}
 \acro{UHF}{Ultra High Frequency}
 \acro{UID}{Unique Identifier} 
 \acro{USRP}{Universal Software Radio Peripheral}
 \acro{USRP2}{Universal Software Radio Peripheral (version 2)}
 \acro{USB}{Universal Serial Bus} 
 \acro{VHF}{Very High Frequency}
 \acro{VLF}{Very Low Frequency}
 \acro{VHDL}{VHSIC (Very High Speed Integrated Circuit) Hardware Description Language}
 \acro{VR}{Voltage Regulator}
 \acro{WLAN}{Wireless Local Area Network}
 \acro{XOR}{Exclusive OR}
 \acro{IR}{Intermediate Representation}
 \acro{OCD}{On-Chip Debug}
 \acro{OS}{Operating System}
 \acro{VM}{Virtual Machine}
 \acro{DVFS}{Dynamic Voltage and Frequency Scaling}
 \acro{MMIO}{Memory-Mapped I/O}
 \acro{TZASC}{TrustZone Address Space Controller}
 \acro{CC}{Confidential Computing}
 \acro{ML}{Machine Learning}
 \acro{DT}{Decision Tree}
 \acro{NN}{Neural Network}
 \acro{DNN}{Deep Neural Network}
 \acro{MLaaS}{Machine Learning as a Service}
 \acro{LLMaaS}{LLM as a Service}
 \acro{XAI}{Explainable Artificial Intelligence}
 \acro{GBDT}{Gradient Boosted Decision Tree}
 \acro{LLM}{Large Language Model}
 \acro{KV}{key--value}
 \acro{FIA}{Fault Injection Attack}
 \acro{SCA}{Side-Channel Attack}
 \acro{BFA}{Bit-flip Attack}
\end{acronym}

\acused{AES}
\acused{CRC}
\acused{DES}
\acused{EEPROM}
\acused{RSA}
\acused{USB}
\acused{SHA-1}
\acused{IC}
\acused{CPU}
\acused{USB}
\acused{DRAM}

\maketitle

\begin{abstract}
Large language models (LLMs) are increasingly deployed for sensitive services, e.g.\ for medical advice. Modern LLM serving frameworks accelerate inference by retaining previously computed states in a key–value (KV) cache. 
Paged serving frameworks such as vLLM organize this cache through block tables analogous to OS page tables, which map each request's logical KV blocks to physical KV blocks, forming a software-managed address translation layer.
This mapping is security-critical serving metadata that determines which cached states each request accesses in multi-tenant serving.
Although fault attacks such as Rowhammer have exploited OS page tables to redirect memory accesses, prior fault attacks on LLMs have largely targeted the \textit{data plane}, \ie model weights or KV-cache contents, leaving KV-cache mapping unexplored.

We present \sys (\ie Block Hijack), the first KV cache address-translation fault attack in LLM inference. 
By corrupting block-table entries, \sys redirects a request's KV cache reads to blocks assigned to another tenant's request without modifying model parameters, prompts, or cached KV states. 
It achieves both confidentiality violation (victim information appears in attacker responses) and integrity violation (attacker-supplied context steers victim answers). 
We study paged KV cache mappings in vLLM and formally characterize when redirect targets are predictable. 
Across seven models spanning the Qwen, Llama, Gemma, and Phi families, a faulted mapping exposes victim context. 
For the integrity attack, we examine how topic proximity, intra-topic divergence, and model behavior shape redirected responses.
On a 248-scene medical benchmark, the integrity attack causes up to 41.1\% of responses to recommend less urgent care, while still appearing consistent with the victim’s question. 
We demonstrate physical feasibility of \sys with Rowhammer and subsequently evaluate assignment validation as a defense. 
Following responsible disclosure, vLLM and Sonar confirmed the vulnerability and proposed patches that we reviewed and validated.

\end{abstract}

\section{Introduction}
\label{sec:introduction}

\acp{LLM} are rapidly deployed in the real world for their exceptional generative capabilities in terms of language understanding and generation tasks.
Recent systems use them for software development (\eg OpenHands~\cite{wang2025openhands}), clinical reasoning (\eg AMIE~\cite{lievin2026conversational}), and scientific discovery (\eg Co-Scientist~\cite{gottweis2026coscientist}).
Serving \ac{LLM} models are increasingly expensive as model size, context length, and request concurrency grow. Autoregressive inference acceleration commonly uses the \ac{KV} cache management that stores the attention keys and values of previously processed tokens, avoiding their recomputation at each decoding step. 
However, the cache grows with the number and length of requests, while variable request sizes can cause memory fragmentation and over-reservation. 
Serving frameworks therefore employ delicate \ac{KV} cache management mechanisms to improve memory utilization and throughput.
The mainstream framework/engine vLLM~\cite{kwon2025vllm} uses PagedAttention~\cite{kwon2023pagedattention} to manage the \ac{KV} cache in fixed-size blocks. The vLLM-derived Sonar engine~\cite{dphnai2026sonar} follows the same block-based design.
SGLang~\cite{zheng2024sglang}, in contrast, uses RadixAttention to organize and reuse cached prefixes.

Alongside these efficiency improvements, prior work has revealed security risks in \ac{LLM} serving~\cite{weiss2024your,wu2025know,wu2026cacheme,zheng2024inputsnatch,song2025early,liu2024formalizing}.
Side-channel attacks infer prompts or responses from token lengths, cache-dependent timing, and scheduling behavior~\cite{weiss2024your,wu2025know,zheng2024inputsnatch,song2025early}, while manipulation attacks influence inference through injected instructions or inputs crafted to induce incorrect cache hits~\cite{wu2026cacheme,liu2024formalizing}.
Serving frameworks like vLLM have since adopted corresponding countermeasures~\cite{wu2026cacheme}.
Fault attacks, however, introduce a distinct path through direct corruption of inference state.

Recently, \acp{LLM} are increasingly accessed through \ac{LLMaaS} APIs to provide service to a volume of tenants, backed by cloud-hosted CPU and GPU infrastructure, where mutually untrusted tenants may share the same physical hardware.
This shared deployment setting makes \acp{FIA} a concern for both tenant confidentiality and integrity of \ac{LLM} inference.
\acp{FIA} have previously been used against \acp{NN} to degrade accuracy~\cite{hong2019terminal} or extract model parameters~\cite{rakin2022deepsteal}.
Recent studies have extended these attacks to \acp{LLM} inference by targeting model weights~\cite{coalson2025prisonbreak,guo2026tfl} or \ac{KV}
cache contents~\cite{yamamoto2026bitflip,alnahian2026cachetrap} in the \textit{data plane}.
Despite differing assumptions and knowledge requirements, these attacks corrupt values used directly in inference, \ie model parameters or cached key/value vectors.

Outside \ac{ML} centered FIA, prior \ac{OS} centered \ac{FIA} exploits have shown that metadata managing memory access is an effective fault target since a localized fault can redirect otherwise legitimate reads or writes across OS isolated boundaries.
These attacks exploit this property by corrupting CPU page-table structures~\cite{vanderveen2016drammer,xiao2016one}, redirecting nested pointers~\cite{tobah2024gogo}, or forging GPU page-table mappings~\cite{wan2026geforge}.

Our insight is that paged \ac{KV} cache mappings serve a similar role in \ac{LLM} serving by determining which cached state each request reads during inference. We note that all existing security threat audits focus on the raw data plane and overlook this metadata layer. Therefore, we ask the following research question: 

\textit{Is the \ac{KV} cache mapping vulnerable to \ac{FIA}? If so, what are the consequences and to what extent?}

\noindent\textbf{Our Contribution.} Through a systematic analysis of this unexplored attack surface, we present \sys, the first \ac{FIA} targeting the address translation mapping that associates each request with its paged \ac{KV} cache state in \ac{LLM} serving. Specifically,
we analyze this attack surface in mainstream \ac{LLM} serving frameworks.
The core observation is that a corrupted mapping can still appear \textit{valid} but redirect/hijack a request's \ac{KV} cache access to cached state belonging to another tenant's request.
Interestingly, depending on whose (either the attacker's or the victim's) mapping metadata is faulted, two distinct attack objectives are rendered:
\begin{itemize}[noitemsep,topsep=0pt]
    \item \textbf{Confidentiality violation.}  The fault changes a mapping in the block table for the \textit{attacker's request}, redirecting its \ac{KV} cache reads to the victim's cached state. The response returned to the attacker can then reveal sensitive information of the victim's context. We call this \emph{Context Disclosure} attack (\siphon). 
    \item \textbf{Integrity violation.} In contrast, the fault changes a mapping in the block table for the \textit{victim's request}, redirecting its \ac{KV} cache reads to cached state derived from the attacker's request.
    This allows attacker-supplied context to steer the victim's response. We call this \emph{Response Alteration} attack (\graft).
\end{itemize}

\noindent\textbf{Technical and security implications of \sys.}
Compared with prior \acp{FIA} that target model weights or \ac{KV} values~\cite{hong2019terminal,yamamoto2026bitflip,alnahian2026cachetrap}, \sys targets the mapping metadata itself, with three implications for \ac{LLM} security and privacy:
(i) Prior work needs to rely on a delicate, vulnerable bit among a huge number of raw bits (\ie of weights or \ac{KV} cache) to be faulted to change the model behavior indeed. Randomly faulting a raw bit will not induce harmful behaviour, since trained \acp{NN} and \acp{LLM} tolerate some blind perturbations~\cite {rakin2019bitflip,yao2020deephammer}, attributing to high sparsity.
The block table instead plays a role analogous to an \ac{OS} page table: changing one entry directly redirects access; each bit of this metadata is critical due to zero sparsity, and a fault is therefore likely to induce unexpected and observable consequences.
(ii) Not every corrupted block-table entry produces a useful attack outcome to the attacker's expectations.
We thus distinguish \emph{targeted} and \emph{untargeted} attacks in \sys, formalize the conditions under which the target remains predictable (\ie controlled and preset by the attacker), and identify when prediction becomes probabilistic and untargeted.
In the targeted case, \sys redirects inference to the intended cached state.
In the untargeted case, the corrupted entry may instead reach a block belonging to an arbitrary resident request.
The mapping space is much smaller than the spaces of model weights or \ac{KV} values, and therefore an untargeted fault may still produce a harmful outcome.
(iii) The attacker-supplied context can shape how \graft alters the victim's response, especially in targeted attack.
In our medical benchmark, this redirection can suppress the urgency of recommended care while keeping the response consistent with the victim's question.
Such integrity violations can have immediate safety consequences in high-stakes applications. 
On October 29, 2026, a news report described a lawsuit alleging harm associated with a ChatGPT-generated recommendation to alter medication dosage, highlighting the risks~\cite{californiacourts2026chatgpt}.
We therefore identify and evaluate three factors that affect whether a redirected response shifts in severity while still being judged consistent with the victim's original question: topic proximity, intra-topic divergence, and victim-model behavior.
Together, these properties underscore \ac{KV} cache mappings as a fault-sensitive target under both targeted and untargeted attacks.

Our medical benchmark-focused experiments demonstrate that a single fault can achieve either confidentiality violation (\siphon) or integrity violation (\graft). 
Confidentiality demonstrations reveal victim context in the attacker's response across seven \ac{LLM} models from four families.
For integrity, we illustrate the effect in an \ac{LLM} medical service across 248 scenarios derived from patient messages in the ChatDoctor-HealthCareMagic-100k dataset~\cite{li2023chatdoctor}.
Across seven served \ac{LLM} models and five \ac{LLM} judges, a covert downgrade occurs when the corrupted response recommends a lower level of care than the uncorrupted response while still being judged to fit the user's question.
The \emph{covert downgrade rate} ranges from 5.2\% to 41.1\%.
We further implement \acp{PoC} for both \siphon and \graft through physical Rowhammer using the modular framework SWAGE~\cite{boy2026slashdsa}.
To evaluate feasibility across frameworks, we reproduce both attack outcomes in SGLang and Sonar using software-emulated experiments.

Our main contributions are summarized as follows:
\begin{itemize}[leftmargin=1.2em,topsep=0.2em,itemsep=0.2em]
    \item We identify previously overlooked paged \ac{KV} cache mappings as a distinct attack surface in \ac{LLM} serving infrastructure and present \sys, a fault attack that enables cross-user information leakage and integrity violations.
    \item We systematically analyze the conditions required for \ac{KV} cache redirection and analyze when a specific victim context can be targeted. We further identify and quantify the factors that shape the resulting confidentiality and integrity outcomes.
    \item We primarily evaluate \sys across multiple \ac{LLM} models in the widely used vLLM and further validate both attack outcomes in SGLang and Sonar serving engines.
    We implement \sys against vLLM in physical Rowhammer experiments using the modular framework SWAGE~\cite{boy2026slashdsa}.
    \item We analyze defenses against \sys and review and validate assignment-validation patches proposed by vLLM and Sonar (\ie a similar fix outlined by our disclosure), confirming the effectiveness while highlighting the security--performance trade-off on the inference path.
\end{itemize}

\noindent\textbf{Responsible Disclosure.} 
We responsibly disclosed our findings to the maintainers of vLLM and SGLang on 15 September 2026 and to Sonar on 24 September 2026.
The vLLM team confirmed our finding as a vulnerability on 22 September 2026, and the Sonar maintainer independently reproduced the issue on 10 November 2026.
We provided technical feedback and helped review and test the proposed fixes.

\section{Background and Related Work}
This section introduces paged \ac{KV} cache management and the serving frameworks considered in this work, then reviews fault attacks and prior attacks on \ac{LLM}.

\subsection{\ac{LLM} Serving Frameworks and \acs{KV} Cache Management}

\noindent\textbf{PagedAttention.}
Inspired by virtual memory and paging in the traditional \ac{OS}, PagedAttention~\cite{kwon2023pagedattention} splits the \ac{KV} cache of an \ac{LLM} into non-contiguous, fixed-size physical blocks (or ``page''). 
It partitions each request's token positions into \emph{logical \ac{KV} blocks} and stores the corresponding \ac{KV} states in \emph{physical \ac{KV} blocks}.
The allocator manages these physical \ac{KV} blocks in a \emph{block pool}, and each block is identified by a fixed integer \emph{block id}.
Every active request occupies one row in the \emph{block table}.
Entry $i$ in that row stores the \emph{block id} of the physical \ac{KV} block corresponding to logical \ac{KV} block $i$.
The \emph{block table} is therefore the serving metadata that implements the logical-to-physical mapping.

\begin{figure*}[htbp]
\centering
\begin{tikzpicture}[font=\scriptsize,
  cell/.style  = {kvblock,minimum width=0.68cm,minimum height=0.46cm,font=\scriptsize},
  pcell/.style = {kvblock,minimum width=0.58cm,minimum height=0.48cm,font=\scriptsize},
  scell/.style = {kvblock,minimum width=0.58cm,minimum height=0.42cm,font=\scriptsize},
  wide/.style  = {kvblock,fill=white,minimum height=0.46cm,inner xsep=4pt,font=\scriptsize},
  note/.style  = {font=\scriptsize,text=black!85,align=left},
  hdr/.style   = {font=\scriptsize\bfseries},
  flow/.style  = {->,>=stealth,line width=0.7pt,draw=black!60},
  lens/.style  = {draw=black!65,line width=0.7pt,densely dashed},
  stage/.style = {draw=black!30,rounded corners=3pt,dashed,line width=0.5pt}]

\node[hdr,anchor=west,xshift=-5pt] at (0,0.72) {\circled{1} Split each request into logical \acs{KV} blocks};
\node[note,anchor=center] at (2.75,0.15) {block size $B=16$};

\node[wide,kvvictim] (ra) at (2.75,-0.38) {\textbf{Request A}: 32 tokens};
\draw[flow] (ra.south) -- ++(0,-0.20);
\node[cell,kvvictim,minimum width=1.55cm] (la0) at (1.88,-1.12) {0--15};
\node[cell,kvvictim,line width=0.9pt,minimum width=1.55cm] (la1) at (3.62,-1.12) {16--31};

\node[wide,kvattacker] (rb) at (2.75,-1.98) {\textbf{Request B}: 32 tokens};
\draw[flow] (rb.south) -- ++(0,-0.20);
\node[cell,kvattacker,minimum width=1.55cm] (lb0) at (1.88,-2.72) {0--15};
\node[cell,kvattacker,minimum width=1.55cm] (lb1) at (3.62,-2.72) {16--31};

\node[note,anchor=center] at (2.75,-3.26)
  {Each request contains two logical \acs{KV} blocks.};
\begin{scope}[on background layer]
  \node[stage,fit={(-0.18,0.44) (5.72,-3.50)}] (s1) {};
\end{scope}

\def\bx{6.75}
\node[hdr,anchor=west,xshift=-5pt] at (\bx,0.72) {\circled{2} Locate \acs{KV} cache blocks through the block table};
\begin{scope}[xshift=-0.85cm]

\draw[black!55,line width=0.6pt,rounded corners=2pt]
      ({\bx+4.02},0.06) rectangle ({\bx+6.28},-1.34);
\node[note,font=\scriptsize\itshape,anchor=south west,inner sep=1pt]
      at ({\bx+4.02},0.08) {block table};

\node[note,anchor=east] at ({\bx+3.96},-0.14) {logical \acs{KV} blk};
\node[note] at ({\bx+4.45},-0.14) {0};
\node[note] at ({\bx+5.15},-0.14) {1};
\node[note] at ({\bx+5.85},-0.14) {$\cdots$};

\node[note,anchor=east] at ({\bx+3.96},-0.56) {\textbf{A} \ $T_A$};
\node[cell,kvvictim] (A0) at ({\bx+4.45},-0.56) {1};
\node[cell,kvvictim,line width=0.9pt] (A1) at ({\bx+5.15},-0.56) {2};
\node[cell,kvvictim] (Arest) at ({\bx+5.85},-0.56) {$\cdots$};

\node[note,anchor=east] at ({\bx+3.96},-1.06) {\textbf{B} \ $T_B$};
\node[cell,kvattacker] (B0) at ({\bx+4.45},-1.06) {3};
\node[cell,kvattacker] (B1) at ({\bx+5.15},-1.06) {4};
\node[cell,kvattacker] (Brest) at ({\bx+5.85},-1.06) {$\cdots$};

\draw[black!60,line width=0.5pt]
  ({\bx+6.42},-0.79) -- ({\bx+6.56},-0.79) -- ({\bx+6.56},-0.33) -- ({\bx+6.42},-0.33);
\draw[black!60,line width=0.5pt]
  ({\bx+6.42},-1.29) -- ({\bx+6.56},-1.29) -- ({\bx+6.56},-0.83) -- ({\bx+6.42},-0.83);
\node[note,anchor=west] at ({\bx+6.66},-0.56) {\emph{row} of request~A};
\node[note,anchor=west] at ({\bx+6.66},-1.06) {\emph{row} of request~B};

\node[note,anchor=center,font=\scriptsize\bfseries] at ({\bx+5.10},-1.68)
  {Physical \acs{KV} blocks in the block pool};
\foreach \i/\sty in {0/kvneutral,1/kvvictim,2/kvvictim,3/kvattacker,
                     4/kvattacker,5/kvneutral,6/kvneutral}{
  \node[pcell,\sty] (p\i) at ({\bx+3.20+0.62*\i},-2.12) {\i};}
\node[pcell,kvvictim,line width=0.9pt] at ({\bx+3.20+0.62*2},-2.12) {2};
\node[note,anchor=west] at ({\bx+7.28},-2.12) {$\cdots$};
\node[note,anchor=north] at ([yshift=2pt]p0.south) {\emph{reserved}};

\node[note,anchor=center,fill=white,inner sep=1.5pt] at ({\bx+5.18},-2.80)
  {\acs{KV} block (id~2): token positions};
\node[scell,kvvictim] (z0) at ({\bx+3.18},-3.22) {16};
\node[scell,kvvictim] (z1) at ({\bx+3.88},-3.22) {17};
\node[note] at ({\bx+4.58},-3.22) {$\cdots$};
\node[scell,kvvictim] (z4) at ({\bx+5.28},-3.22) {20};
\node[note] at ({\bx+5.98},-3.22) {$\cdots$};
\node[scell,kvvictim] (z14) at ({\bx+6.68},-3.22) {30};
\node[scell,kvvictim] (z15) at ({\bx+7.38},-3.22) {31};
\begin{scope}[on background layer]
\draw[lens] (p2.south west) -- (z0.north west);
\draw[lens] (p2.south east) -- (z15.north east);
\end{scope}
\end{scope}
\begin{scope}[on background layer]
  \node[stage,fit={({\bx-0.18},0.44) ({\bx+8.75},-3.50)}] (s2) {};
\end{scope}
\draw[flow] (s1.east) -- (s2.west);

\end{tikzpicture}
\caption{\textbf{Illustrative paged \acs{KV} cache mapping in vLLM.}
With block size $B=16$, each 32-token request contains two logical \acs{KV} blocks.
Each request occupies one block-table row.
The entry at logical index $i$ stores the id of the corresponding physical \acs{KV} block.
The physical \acs{KV} block with id~2 is expanded to show its 16 token slots.} 
\label{fig:vllm-pipeline}
\end{figure*}
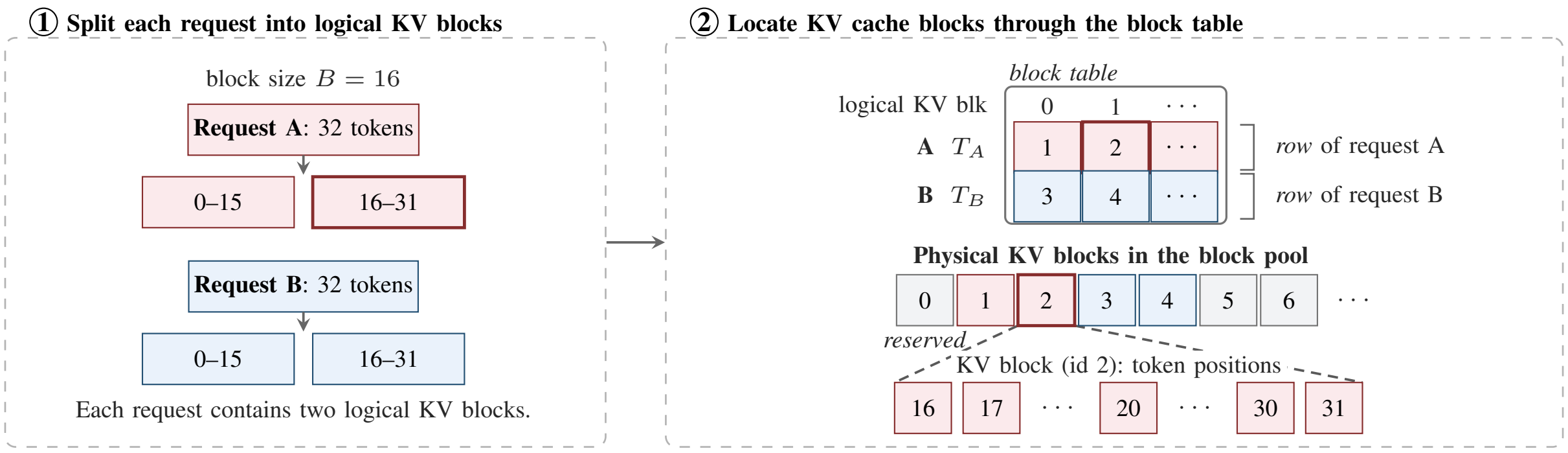

\noindent\textbf{vLLM.}
vLLM implements PagedAttention~\cite{kwon2023pagedattention} for \ac{KV} cache management, combining block-table mappings with a dynamically managed pool of physical \ac{KV} blocks.
We illustrate these structures and their relationship with a toy example in \Cref{fig:vllm-pipeline}.
vLLM ships two model runners that differ in how they maintain the block table. 
Model Runner~V1 (MRV1) stores it in CPU host memory. 
With the GPU backend, it copies the active batch's rows to GPU memory before each forward pass.
Under the CPU backend, the table remains in host memory and is used there.
Model Runner~V2 (MRV2), the default for dense models since v0.25, holds it directly in GPU memory under the GPU backend and in CPU host memory under the CPU backend~\footnote{We validated GPU-backed MRV2 and MRV1 against vLLM commits \href{https://github.com/vllm-project/vllm/blob/0ecc284790e5403f74b899524ef82ecb69f83cb3/vllm/v1/worker/gpu/block\_table.py\#L47-L54}{\texttt{0ecc284}} and \href{https://github.com/vllm-project/vllm/blob/95582868efd4db0b120e3640bbc61dcfce20d59f/vllm/v1/worker/block\_table.py\#L166-L167}{\texttt{9558286}}, respectively. 
We additionally validated CPU-backed MRV2 using vLLM \texttt{v0.30.0+cpu} (\texttt{gced6857af}).}.

\noindent\textbf{SGLang and Sonar.}
SGLang~\cite{zheng2024sglang} uses RadixAttention to organize prefix matching, reuse, and eviction while storing \ac{KV} states in a paged layout.
Its request-to-token mapping determines which \ac{KV} slots a request accesses.
Sonar~\cite{dphnai2026sonar} is a vLLM-based inference engine that likewise provides paged \ac{KV} cache management.

We use vLLM as our primary platform.
\Cref{sec:other_platforms} extends the analysis to SGLang and Sonar.

\subsection{Fault Attacks}
\noindent\textbf{Rowhammer attacks.}
Rowhammer repeatedly activates a \ac{DRAM} row to induce bit flips in physically adjacent rows~\cite{kim2014flipping}.
On CPUs, prior work has developed address-mapping and TRR-bypass techniques
and used the resulting faults to cross memory-isolation boundaries, corrupt security-critical pointers, or disclose sensitive data
~\cite{vanderveen2016drammer,tobah2024gogo,razavi2016flip,kwong2020rambleed,pessl2016drama,frigo2020trrespass,jattke2022blacksmith}.
More recently, GPUHammer~\cite{lin2025gpuhammer} demonstrated Rowhammer on discrete-GPU GDDR6,
while GeForge~\cite{wan2026geforge}, GDDRHammer~\cite{hu2026gddrhammer}, and GPUBreach~\cite{lin2026gpubreach} turned GPU faults into page-table
attacks.
GPUThor~\cite{lin2026gputhor} further demonstrated attacks on an ECC-enabled GPU.
Together, these studies show that Rowhammer can corrupt security-critical data and mappings in both CPU and GPU memory.

\noindent\textbf{Other fault-injection methods.}
Beyond Rowhammer, hardware faults can be induced through clock or voltage glitching, electromagnetic injection, and optical injection, which offer different spatial and temporal precision~\cite{chen2021voltpillager,horvath2025barracuda,muench2025transparency}.
Clock and voltage glitching disturb circuit timing or supply conditions during selected operations.
Electromagnetic injection applies a localized pulse to induce transient currents, whereas optical injection uses focused light, typically a laser, to perturb selected circuit elements.

\subsection{Related Work}
Prior work on \ac{LLM} serving attacks can be divided according to whether it observes serving behavior or actively alters the inputs or state used during inference.
We classify attacks as \emph{passive} when they infer information without changing the victim's input, model, or cached state, and as \emph{active} when they manipulate inputs or \ac{KV} cache.
This section summarizes the most relevant works.

\subsubsection{Passive attacks}
Passive attacks recover the victim's sensitive information from observable behavior in shared \ac{LLM} serving.
Weiss \etal~\cite{weiss2024your} infer per-token lengths from the packet sizes of encrypted streaming responses, then use fine-tuned \acp{LLM} to reconstruct the plaintext.
Zheng \etal~\cite{zheng2024inputsnatch} and Song \etal~\cite{song2025early} exploit response-time differences between cache hits and misses in prefix/\ac{KV} and semantic caches to infer private user inputs and system prompts.   
Wu \etal~\cite{wu2025know} instead exploit longest-prefix-match scheduling: a candidate request that extends the victim's cached prefix receives higher scheduling priority, and its return order reveals whether the guessed token matches.
Passive attacks therefore alter neither the victim's model input nor the cached state maintained by the serving engine.

\subsubsection{Active attacks}
Active attacks, by contrast, influence inference by manipulating its inputs or internal state.
We classify them as application-layer and fault-based attacks according to how the modification is introduced, and compare them in \Cref{tab:related-work-summary}.

\begin{table}[htbp]
    \centering
    \scriptsize
    \setlength{\tabcolsep}{3.2pt}
    \renewcommand{\arraystretch}{1.12}
    \newcommand{\yesmark}{\ding{51}}
    \newcommand{\nomark}{\ding{55}}

    \caption{\textbf{Comparison of active attacks on \ac{LLM} serving.}
    C and I denote confidentiality and integrity violations, respectively.}
    \label{tab:related-work-summary}

    \begin{adjustbox}{width=\linewidth}
    \begin{tabular}{llccccc}
        \toprule

        \multirow{2}{*}{\textbf{Work}} &
        \multirow{2}{*}{\textbf{Method}} &
        \multicolumn{3}{c}{\textbf{Target}} &
        \multicolumn{2}{c}{\textbf{Security Impact}} \\

        \cmidrule(lr){3-5}
        \cmidrule(lr){6-7}

        & &
        \makecell{\textbf{Input/}\\\textbf{Lookup}} &
        \makecell{\textbf{Weights/}\\\textbf{\acs{KV} Values}} &
        \makecell{\textbf{\acs{KV}}\\\textbf{Mapping}} &
        \textbf{C} &
        \textbf{I} \\

        \midrule
        \multicolumn{7}{l}{\textit{Application-layer manipulation}} \\

        Prompt injection~\cite{liu2024formalizing} &
        API input &
        \yesmark & \nomark & \nomark &
        \nomark & \yesmark \\

        Cache Me~\cite{wu2026cacheme} &
        \makecell[l]{API cache\\ collision} &
        \yesmark & \nomark & \nomark &
        \nomark & \yesmark \\

        \addlinespace[0.35em]
        \multicolumn{7}{l}{\textit{Fault-based manipulation}} \\

        Yamamoto and Matsuura~\cite{yamamoto2026bitflip} &
        Software \ac{FIA} &
        \nomark & \yesmark & \nomark &
        \nomark & \yesmark \\

        CacheTrap~\cite{alnahian2026cachetrap} &
        Rowhammer &
        \nomark & \yesmark & \nomark &
        \nomark & \yesmark \\

        PrisonBreak~\cite{coalson2025prisonbreak} &
        Rowhammer &
        \nomark & \yesmark & \nomark &
        \nomark & \yesmark \\

        TFL~\cite{guo2026tfl} &
        Rowhammer &
        \nomark & \yesmark & \nomark &
        \nomark & \yesmark \\

        \addlinespace[0.2em]
        \rowcolor{gray!15}
        \textbf{\sys{} (this work)} &
        Rowhammer &
        \nomark & \nomark & \yesmark &
        \yesmark & \yesmark \\

        \bottomrule
    \end{tabular}
    \end{adjustbox}
\end{table}

\noindent\textbf{Application-layer manipulation.}
Prompt injection places attacker-controlled instructions in external data processed by an \ac{LLM}-integrated application, causing the model to perform an attacker-chosen task~\cite{liu2024formalizing}.
PromptLocate~\cite{jia2026promptlocate} localizes the injected content to support detection and recovery.
Wu \etal~\cite{wu2026cacheme} exploit collisions in prefix, semantic, and multimodal cache lookup to make later requests reuse attacker-prepared cached state, enabling cache poisoning and content-moderation bypass.
Application-layer attacks pass through explicit input and cache-lookup interfaces, enabling direct countermeasures such as prompt localization, cache salting, and collision-resistant hashing. \ac{FIA}, however, bypasses these interfaces and corrupts inference data, serving code, or metadata, shifting defense from interface hardening to protecting and validating internal serving state.

\noindent\textbf{Fault-based manipulation.}
Most prior work targets the \emph{data plane}, \ie model weights or cached \ac{KV} values.
Yamamoto and Matsuura~\cite{yamamoto2026bitflip} use ideal software injection to study a single BF16 bit flip in the value tensor of a shared prefix block in vLLM. This design enables broad exploration of \ac{KV} cache bit flips, while their realization under hardware-specific Rowhammer constraints, including the availability and direction of physical flips, is not evaluated.
CacheTrap~\cite{alnahian2026cachetrap} identifies bits in cached \ac{KV} values that trigger attacker-chosen model behavior and separately reproduces GPUHammer's~\cite{lin2025gpuhammer} bit-flip benchmark.
For model weights, PrisonBreak~\cite{coalson2025prisonbreak} uses targeted bit flips to jailbreak aligned \acp{LLM}, whereas TFL~\cite{guo2026tfl} steers selected prompts toward attacker-specified answers.
Both works demonstrate physical Rowhammer against model weights, on GPU and CPU memory, respectively.
Together, these attacks corrupt numerical values used directly in inference.
In contrast, \sys corrupts a block-table entry, the serving metadata that stores the block id assigned to one logical \ac{KV} block.

\section{System Model and Threat Model}
\label{sec:overview}
This section introduces the multi-tenant \ac{LLM} and gives a clear system model and threat model we considered.

\subsection{Multi-Tenant \acs{LLM}}
Following the standard serving model used by recent work~\cite{weiss2024your,wu2025know,wu2026cacheme,zheng2024inputsnatch,song2025early}, we consider a hosted \ac{LLM} service with different users. 
A user submits a request whose model input is a variable-length token sequence:
\begin{equation}
r \;=\; \left(t_0,\ldots,t_{N_t-1}\right)
\end{equation}
Depending on the service configuration, $r$ may contain both user-provided tokens and tokens prepended by the service.
A \textit{chat template} supplied with the model adds role markers and control tokens, while the deployment may prepend fixed system instructions \cite{lyu2024keeping}. 
These components are shared across users of the same service, leaving the user-provided content as the main source of variation between requests. 

Inference proceeds through two phases: prefill processes the input sequence, and decode generates output tokens autoregressively.
The newly computed key and value states are written to the physical \ac{KV} blocks selected through the request's mapping, attention then reads the cached states within the layer's attention window.

Each logical \ac{KV} block spans $B$ token positions.
For position $p$, define
\begin{equation}
\mathrm{blk}(p)=\left\lfloor p/B \right\rfloor,\qquad
\mathrm{off}\left(p\right)=p\bmod B.
\end{equation}
Here, $\mathrm{blk}(p)$ identifies the logical \ac{KV} block containing $p$, and $\mathrm{off}(p)$ gives its offset within that block.
Let $T_u$ denote the block-table row of request $u$.
The entry $T_u[i]$ stores the block id of the physical \ac{KV} block assigned to logical \ac{KV} block $i$.
Let $K_b$ and $V_b$ denote the sequences of key and value vectors stored in physical \ac{KV} block $b$.
Because layer and \ac{KV} head indices do not affect the block-table lookup, we omit them from the notation.
The cached states at position $p$ are therefore
\begin{equation}
\begin{aligned}
b &= T_u[\mathrm{blk}(p)],\\
(k_p,v_p) &= \bigl(K_b[\mathrm{off}(p)],V_b[\mathrm{off}(p)]\bigr).
\end{aligned}
\label{eq:Kread}
\end{equation}

In vLLM, the row $T_u$ is an \texttt{int32} array of block ids.
For logical \ac{KV} block $i$, both \ac{KV} reads and writes use the id stored in $T_u[i]$.

\subsection{System Model}
We consider two parties: users and an \ac{LLM} service provider.
The provider operates a serving engine comprising a scheduler and one or more workers.
Users submit requests to the service, the scheduler selects queued requests for execution, and the workers perform inference.
The provider then returns the generated outputs to the users.

\noindent\textbf{Users.} 
We assume $N$ users utilize the \ac{LLM} service, with each dispatching its request (denoted by $r$) to the provider.
Each request $r$ consists of $N_t$ tokens, \ie $(t_0,\ldots,t_{N_t-1})$, where $N_t$ varies.
After being processed by the inference workflow, each user receives an output comprising $j$ tokens $(t_{N_t},\cdots, t_{N_t+j-1})$, where $j$ varies and can be specified through user-defined configurations in the serving framework (\eg $\mathtt{max\_tokens}$ in vLLM).

\noindent\textbf{\ac{LLM} service provider.}
The provider operates the serving engine and maintains the model and
\ac{KV} cache resources used for inference.
The \emph{scheduler} selects incoming requests from the queue and dispatches
them to the workers.
Each \emph{worker} executes the deployed \ac{LLM} using the assigned CPU or
GPU resources.

Following prior work~\cite{wu2025know,wu2026cacheme,zheng2024inputsnatch}, we consider the default \emph{First-Come-First-Served} (FCFS) scheduling and continuous batching. 
Prefix caching allows a request to reuse \ac{KV} blocks previously computed for a matching token prefix.
It does not affect the redirect in a fresh \ac{LLM} serving instance. In a noisy state, it matters when predicting the target block ids (see \Cref{subsec:id-prediction}).

\subsection{Threat Model}

\label{sec:threat-model}

We consider a multi-tenant instance that batches requests from different users for the same model. Without loss of generality, we refer to the victim user as Bob and the attacker as Mallory. 
Bob's query contains a sensitive predicate $s_B$, such as a medical diagnosis. 
Mallory's goal is to make her own response reveal information about $s_B$ in \siphon, or to steer Bob's response toward Mallory's content in \graft.

\noindent\textbf{Attacker capabilities.}
In general, Mallory can submit ordinary API requests and observe the responses to her own requests.
Beyond this API access, Mallory can induce a fault that changes the block id stored in a selected block-table entry.
We use Rowhammer as an illustrative fault primitive for \sys to induce this fault and adopt a standard Rowhammer-based fault model~\cite{vanderveen2016drammer,razavi2016flip,kwong2020rambleed,lin2025gpuhammer,wan2026geforge,lin2026gputhor,hu2026gddrhammer,lin2026gpubreach}: Mallory is an unprivileged attacker with software-only access who shares a physical machine with the victim.
We therefore assume that the memory containing this entry is susceptible to Rowhammer-induced bit flips.
Such co-residence can arise on shared servers, in multi-tenant cloud infrastructure, and in \ac{LLMaaS} deployments.
We assume that the \ac{OS} and hypervisor operate correctly and enforce process isolation and memory protection.
Mallory therefore cannot directly read or write the service's address space, modify its model files, or control the victim's inference requests.
The required Rowhammer primitive depends on where the block table is stored.
MRV1 and CPU-backed MRV2 require CPU Rowhammer~\cite{vanderveen2016drammer,razavi2016flip,kwong2020rambleed,boy2026slashdsa}, whereas GPU-backed MRV2 requires GPU Rowhammer~\cite{lin2025gpuhammer,wan2026geforge,lin2026gputhor,hu2026gddrhammer,lin2026gpubreach}.

\noindent\textbf{Attacker knowledge.}
Mallory knows the public serving configuration, including tokenizer, chat template, and input format used by the service, as well as the relevant settings.
Consistent with prior work~\cite{wu2025know,wu2026cacheme,song2025early,coalson2025prisonbreak}, these assumptions capture open-source or documented deployments in which model artifacts and input interfaces are public and the relevant backend settings are disclosed or known from prior profiling.
She does not observe other users' prompts, outputs, or cached \ac{KV} states.

\section{\sys}
This section presents how \sys redirects \ac{KV} cache reads by corrupting a block-table entry.
We then establish the conditions required for \siphon and \graft and analyze when their target block ids can be predicted across allocator states.

\subsection{\acs{KV} Cache Redirection via Mapping Corruption}
\label{sec:mechanics}

We write $M$ for the attacker and $B$ for the victim.
Let $b_M = T_M[\idx_M]$ denote the block id stored at logical index $\idx_M$ of Mallory's block table row, and $b_B = T_B[\idx_B]$ denote the id of the block holding Bob's sensitive predicate $s_B$.
Both ids are assigned by the allocator.
In \siphon the fault alters Mallory's entry, giving the redirect $T_M[\idx_M]:b_M \;\xrightarrow{\mathrm{fault}}\; b_B$.
For every position $p$ with $\mathrm{blk}(p)=\idx_M$, subsequent \ac{KV} reads return
$K_{b_B}[\mathrm{off}(p)]$ and $V_{b_B}[\mathrm{off}(p)]$
instead of Mallory's own cached states.
In \graft, the corresponding fault
$T_B[\idx_B]:b_B\xrightarrow{\mathrm{fault}}b_M$
redirects Bob's \ac{KV} reads to the physical \ac{KV} block identified by $b_M$.

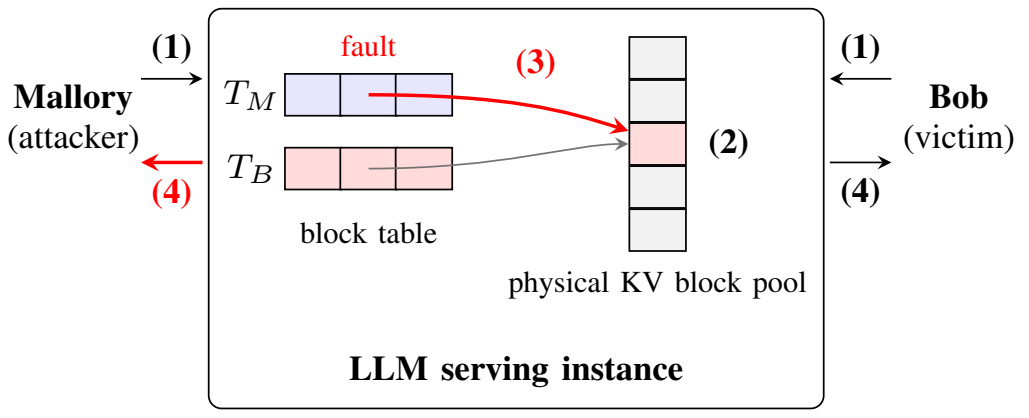
\begin{figure}[t]
\centering
\resizebox{0.8\linewidth}{!}{%
\begin{tikzpicture}[font=\footnotesize,>=stealth,
  c/.style={draw,minimum width=0.46cm,minimum height=0.34cm,inner sep=0pt},
  atk/.style={c,fill=blue!10},
  vic/.style={c,fill=red!14},
  fre/.style={c,fill=black!5},
  n/.style={font=\footnotesize\bfseries}]

\draw[rounded corners=3pt] (-2.5,-1.90) rectangle (2.5,1.35);
\node[anchor=south,font=\footnotesize\bfseries] at (0,-1.85) {\acs{LLM} serving instance};

\node at (-2.15,0.65) {$T_M$};
\node[atk] (m0) at (-1.65,0.65) {};
\node[atk] (m1) at (-1.20,0.65) {};
\node[atk] (m2) at (-0.75,0.65) {};
\node[font=\scriptsize,red,anchor=south] at (-1.20,0.84) {fault};
\node at (-2.15,0.05) {$T_B$};
\node[vic] (v0) at (-1.65,0.05) {};
\node[vic] (v1) at (-1.20,0.05) {};
\node[vic] (v2) at (-0.75,0.05) {};
\node[font=\scriptsize,anchor=north] at (-1.20,-0.26) {block table};

\node[fre] (p0) at (1.15,0.95) {};
\node[fre] (p1) at (1.15,0.60) {};
\node[vic] (p2) at (1.15,0.25) {};
\node[fre] (p3) at (1.15,-0.10) {};
\node[fre] (p4) at (1.15,-0.45) {};
\node[font=\scriptsize,anchor=north,align=center] at (1.15,-0.66) {physical \acs{KV} block pool};
\node[n,anchor=west] at (1.44,0.25) {(2)};

\draw[->,black!55] (v1.center) .. controls (-0.10,0.05) and (0.45,0.25) .. (p2.west);
\draw[->,red,thick] (m1.center) .. controls (-0.05,0.65) and (0.40,0.52) .. ([yshift=-2pt]p2.north west);
\node[n,red] at (0.16,0.90) {(3)};

\node[align=center] (bob) at (3.60,0.45) {\textbf{Bob}\\(victim)};
\node[align=center] (mal) at (-3.60,0.45) {\textbf{Mallory}\\(attacker)};

\draw[->] (3.05,0.78) -- node[n,above]{(1)} (2.55,0.78);
\draw[->] (-3.05,0.78) -- node[n,above]{(1)} (-2.55,0.78);
\draw[->,red,thick] (-2.55,0.10) -- node[n,below,red]{(4)} (-3.05,0.10);
\draw[->] (2.55,0.10) -- node[n,below]{(4)} (3.05,0.10);
\end{tikzpicture}}
\caption{\textbf{Attack overview.} (1) Both users issue the requests; (2) their computed \acs{KV} states are written to the assigned physical \acs{KV} blocks; (3) a fault changes the block id stored in Mallory's entry from $b_M$ to $b_B$, causing the corresponding \acs{KV} reads to access Bob's block; (4) Mallory's reply carries Bob's content. \graft changes Bob's entry from $b_B$ to $b_M$, sourcing Bob's read from Mallory's block.}
\label{fig:overview}
\end{figure}

The redirect yields two distinct outcomes, depending on which request's block-table entry is corrupted.
We provide an overview of the attack in \Cref{fig:overview}.

\begin{itemize}[leftmargin=1.2em,topsep=0.2em,itemsep=0.2em]
\item \textbf{Confidential leakage (\siphon).} 
Bob's prompt contains a sensitive predicate $s_B$ (\eg a diagnosis Mallory did not ask about). 
After the redirect, Mallory's response contains content explaining $s_B$, which is not related to Mallory's own query. 
If combined with side information that narrows the predicate to a small set of candidate identities, such as a patient roster, $s_B$ enables Mallory to perform attribute-based identity inference and link the diagnosis to Bob~\cite{yeom2018privacy,fredrikson2015model}.

\item \textbf{Integrity violation (\graft).} Here the fault is induced in Bob's block table rather than Mallory's. 
Taking \ac{LLM} medical service as example~\cite{li2023chatdoctor}, Bob's prompt describes a condition that requires urgent care, while Mallory's is a benign complaint on the same topic (\eg severe versus mild chest pain). 
The redirect causes the model to answer Bob's question using Mallory's milder context, \ie the response stays on topic but recommends less urgent care than Bob's condition requires.
Understating urgency is more harmful than overstating it because it can lead to inaction rather than an unnecessary consultation.
\end{itemize}

\subsection{Attack Construction}
\label{sec:attack_cons}
We next describe how Mallory realizes the \ac{KV} cache redirection in \sys.

\noindent\textbf{Request co-residency.}
Mallory's and Bob's requests are co-resident, with their physical \ac{KV} blocks allocated from the same block pool and identified within its block-id space.

\noindent\textbf{Redirection window.}
Changing the block id stored in one request's block-table entry redirects the corresponding \ac{KV} reads to the physical \ac{KV} block identified by the new id.
We call this physical \ac{KV} block the \emph{source} and the request whose entry is corrupted the \emph{reader}.
In \siphon, Bob is the source and Mallory is the reader, whereas \graft reverses these roles.
The valid timing window for \sys is the period during which the corrupted mapping produces the intended read redirection.
Formally, it opens once the source block contains its intended \ac{KV} states and the reader has completed its final prefill \ac{KV} write through that entry.
More precisely, the reader need not have completed its entire prefill phase.
In \siphon, for example, only its final prefill write through $T_M[\idx_M]$ should have occurred.

\begin{figure}[t]
\centering
\resizebox{\linewidth}{!}{%
\begin{tikzpicture}[font=\scriptsize,
  bar/.style   = {draw=black!55,line width=0.4pt},
  src/.style   = {bar,fill=victimfill,draw=victimred!75!black},
  rdr/.style   = {bar,fill=attackerfill,draw=attackerblue!75!black},
  blk/.style   = {draw=black!65,line width=0.55pt,fill=neutralfill,rounded corners=2pt},
  blkh/.style  = {draw=black!55,line width=0.7pt,fill=black!5,rounded corners=2pt,dash pattern=on 2.4pt off 1.4pt},
  lab/.style   = {font=\footnotesize,anchor=east},
  ann/.style   = {font=\scriptsize,text=black!80,align=center},
  tick/.style  = {font=\fontsize{6}{7}\selectfont,text=black!80,align=center},
  edge/.style  = {draw=black!70,line width=0.7pt,dash pattern=on 2.4pt off 1.4pt},
  soft/.style  = {draw=black!50,line width=0.55pt,densely dotted}]

\def\wo{4.30}\def\wc{6.70}\def\ov{8.25}\def\po{3.00}
\fill[black!11] (\wo,-0.62) rectangle (\wc,1.20);
\fill[black!4]  (\wc,-0.62) rectangle (\ov,1.20);
\draw[edge] (\wo,-0.62) -- (\wo,1.20);
\draw[edge] (\wc,-0.62) -- (\wc,1.20);
\draw[soft] (\ov,-0.62) -- (\ov,1.20);
\draw[soft] (\po,-0.62) -- (\po,1.20);

\node[lab] at (2.02,0.95) {source request};
\draw[src] (2.20,0.78) rectangle (\po,1.12);   \node[ann] at (2.63,0.93) {prefill};
\draw[src] (\po,0.78) rectangle (5.10,1.12);   \node[ann] at (4.15,0.95) {decode};
\node[ann,anchor=west,text=black!55] at (5.18,0.95) {};

\node[lab] at (2.02,0.35) {source block};
\draw[blk]  (2.20,0.16) rectangle (\po,0.54);  \node[ann] at (2.60,0.35) {written};
\draw[blkh] (\po,0.16) rectangle (\ov,0.54);   \node[ann] at (5.60,0.35) {retains the source \acs{KV} states};

\node[lab] at (2.02,-0.25) {reader request};
\draw[rdr] (2.55,-0.42) rectangle (\wo,-0.08); \node[ann] at (3.42,-0.25) {prefill};
\draw[rdr] (\wo,-0.42) rectangle (\wc,-0.08);  \node[ann] at (5.50,-0.25) {decode};

\node[ann,anchor=south,text=victimred,align=center] at ({(\wo+\wc)/2},1.22)
  {\textbf{timing window}\\mapping remains corrupted};

\node[tick,anchor=north,align=center] at (\wo,-0.68) {\textbf{opens}\\reader's last write};
\node[tick,anchor=north,align=center] at (\wc,-0.68) {\textbf{closes}\\reader's last read};
\node[tick,anchor=north,align=center,text=black!60] at (\po,-0.68) {source block\\populated};
\node[tick,anchor=north,align=center,text=black!60] at (\ov,-0.68) {source block\\overwritten};

\node[tick,anchor=west,align=left] at (1.05,-1.62)
  {\siphon: source $=$ Bob, \ \ \ \ reader $=$ Mallory \ \ \ ($T_M[\idx_M]:b_M\!\to\!b_B$)};
\node[tick,anchor=west,align=left] at (1.05,-1.96)
  {\graft: source $=$ Mallory, reader $=$ Bob \ \ \ \ \ \ \ ($T_B[\idx_B]:b_B\!\to\!b_M$)};

\end{tikzpicture}}
\caption{\textbf{Illustrative window of effective \acs{KV} cache redirection.}
Throughout this window, the mapping entry remains corrupted, so the reader accesses the redirected \acs{KV} block at each decode step.}
\label{fig:window}
\end{figure}
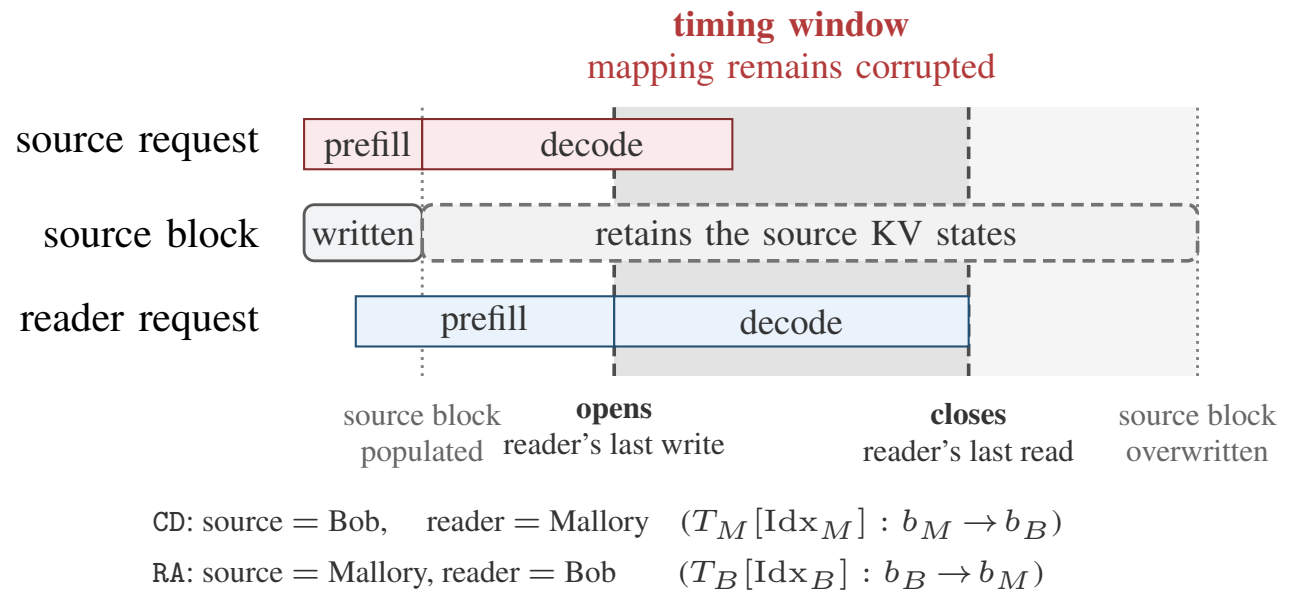

\Cref{fig:window} illustrates this window with a simplified schedule.
In practice, batching and chunked prefill may cause the two prefills to overlap with decoding.
These scheduling differences change the duration of the phases but not the events that bound the redirection window.

\noindent\textbf{Target block id selection.}
In \sys, the required fault changes the block id stored in one block-table entry so that it identifies a different physical \ac{KV} block.
A key prerequisite is therefore how Mallory predicts the block id holding Bob's sensitive information $s_B$ (\ie targeted attack).
We focus on a vLLM service starting from its initial state, in which Mallory knows the chat template and the serving configuration and can thus infer the logical index of $s_B$ (predictions across noisy states are discussed in \Cref{subsec:id-prediction}).

A \emph{same-index} redirect uses Mallory's and Bob's entries at the same logical \ac{KV} block index, \ie $\idx_M=\idx_B$, thereby preserving their token-position alignment in models using rotary position embedding (RoPE)~\cite{su2021roformer}.
A \emph{cross-index} redirect instead selects blocks at different logical block indices, \ie $\idx_M\neq\idx_B$.
It preserves the within-block offsets, while the redirected keys retain Bob's original RoPE positions, producing a constant position shift of $B \times (\idx_B-\idx_M)$.
This shift changes the attention scores but does not prevent the source \ac{KV} states from being read, so the source content can still affect the response (\eg demonstrated example in \Cref{fig:case1-cross}).
Cross-index selection expands the set of candidate block-id transitions, allowing Mallory to exploit a broader range of Rowhammer-induced bit flips.
Mallory can lengthen her request to obtain more candidate logical blocks and insert dummy requests to shift their assigned ids.
These dummy requests are ordinary requests that Mallory keeps active to consume intervening block ids before her selected block is allocated.

\noindent\textbf{Fault primitive.}
Such a fault can for example be induced by voltage/clock glitching~\cite{chen2021voltpillager}, laser fault injection~\cite{muench2025transparency}, \ac{EM} injection~\cite{horvath2025barracuda}, or Rowhammer~\cite{kim2014flipping,frigo2020trrespass,jattke2022blacksmith}. 
These alternatives require different access assumptions: voltage or clock glitching may rely on privileged control or physical access to power-management interfaces, EM injection requires physical proximity, and laser injection typically requires chip decapsulation.
\sys assumes the ability to corrupt the block id stored in a target block-table entry, which our \ac{PoC} realizes using Rowhammer, a software-induced hardware fault primitive.
We demonstrate exploiting \sys using Rowhammer in \Cref{sec:rowhammer_sys}.
More generally, \sys relies on corrupting an entry in the paged \ac{KV} cache mapping to redirect access.
It can therefore be implemented with any attack primitive capable of corrupting the target entry.
We discuss alternative primitives in \Cref{sec:alternative_primitives}.

\subsection{Block ID Prediction across Allocator States}
\label{subsec:id-prediction}
The block ids $b_B=T_B[\idx_B]$ and $b_M=T_M[\idx_M]$
depend on the positions at which the selected blocks are drawn from the allocator's free queue.
In a newly initialized instance, the free queue contains the candidate ids $\{1,\ldots,U\}$ in ascending order, with 0 reserved.
Public information, including the tokenizer, chat template, block size, and fixed input format, allows Mallory to identify the logical block expected to contain $s_B$~\cite{wu2025know,wu2026cacheme}.

We now relax the assumption and consider a serving instance after prior usage, \ie it has processed other requests~\footnote{This analysis follows the allocator policy at commit \texttt{0ecc284}.}.
Let $H$ be the largest id assigned since initialization. Let $\pi_F$ denote previously used ids that are now free, and $\pi_C$ those that are also free but whose cached content remains available for reuse.
Let $F=|\pi_F|$, and let $Z$ count the remaining touched ids, including those currently allocated and those retained in $\pi_C$.
Under allocation, release, and prefix-cache-hit operations, the queue has the form
\begin{equation}
Q=\pi_F\mathbin{\Vert}[H+1,\ldots,U]\mathbin{\Vert}\pi_C
\label{eq:allocator-state}
\end{equation}
The internal orders of $\pi_F$ and $\pi_C$ depend on prior requests, whereas the middle range contains ids that have not yet been assigned.
A hit on a free cached block moves its id from $\pi_C$ into active use and therefore leaves $Z$ unchanged.
Throughout the analysis, $Z$ denotes its value at the start of the controlled allocation sequence.

\noindent\textbf{Allocation positions.}
For the controlled order of Bob, intervening dummy requests, and Mallory, let $\rho_B$ and $\rho_M$ denote the zero-based positions of their selected blocks among the ids drawn from the free queue, after the initial dummy allocations from Mallory, where $0\le\rho_B<\rho_M$.

Suppose Mallory knows $Z$ and a valid upper bound $\bar H\ge H$.
For a service she has profiled over time, Mallory can probably use historical request rates and input/output-length distributions~\cite{zhong2024distserve,wang2025burstgpt} to estimate them.
Algorithm~\ref{alg:id-prediction} computes the number
$D_{\mathrm{pre}}$ of block allocations to place before Bob and predicts the two selected block ids.

\begin{algorithm}[t]
\renewcommand{\algorithmicrequire}{\textbf{Input:}}
\renewcommand{\algorithmicensure}{\textbf{Output:}}
\footnotesize
\caption{Block-id prediction under prior usage}
\label{alg:id-prediction}
\begin{algorithmic}[1]
\Require $U,\bar H,Z,\rho_B,\rho_M$, where
         $0\le\rho_B<\rho_M$
\Ensure $\langle D_{\mathrm{pre}},\hat b_B,\hat b_M\rangle$, or $\bot$
\State $D_{\mathrm{pre}}\gets
       \max(0,\bar H-Z-\rho_B)$
\State $a_B\gets D_{\mathrm{pre}}+\rho_B$
\State $a_M\gets D_{\mathrm{pre}}+\rho_M$
\If{$a_M\ge U-Z$}
    \State \Return $\bot$
\EndIf
\State $\hat b_B\gets Z+a_B+1$
\State $\hat b_M\gets Z+a_M+1$
\State \Return
       $\langle D_{\mathrm{pre}},\hat b_B,\hat b_M\rangle$
\end{algorithmic}
\end{algorithm}

\noindent\textbf{Correctness.}
From \eqref{eq:allocator-state}, the unknown head has length
$F=H-Z$.
The choice of $D_{\mathrm{pre}}$ gives $a_B \ge \bar H-Z \ge H-Z =F$.
The consecutive range occupies queue positions where $[F,F+(U-H))=[F,U-Z)$.
Therefore, when $a_M<U-Z$, both $a_B$ and $a_M$ lie in this range.
An allocation at position $a$ within the range receives id $H+1+(a-F)=H+1+a-(H-Z)=Z+a+1$.
The prediction is therefore independent of the internal orders of $\pi_F$ and $\pi_C$.
A return value of $\bot$ only means that this consecutive-range method does not apply.

\noindent\textbf{Prior usage without prefix caching.}
$\pi_C$ is empty and $Z$ counts the ids that remain allocated.
Prior releases may reorder the ids in $\pi_F$, but do not affect the consecutive range $[H+1,\ldots,U]$.
Given $Z$ and a valid upper bound $\bar H$, Algorithm~\ref{alg:id-prediction} places both selected allocations within this range and predicts their ids, provided that the range is sufficiently large.
If all earlier requests have completed, this reduces to the case $Z=0$.
In this case, Mallory can predict the block ids for Bob despite prior allocator use.

\noindent\textbf{Prior usage with prefix caching.}
With prefix caching enabled, $Z$ also includes free ids retained in $\pi_C$.
A cache hit reuses an id already counted in $Z$ and changes the allocation offsets $\rho_B$ and $\rho_M$.
Exact prediction consequently requires $\bar H$, the initial value of $Z$, and the cache-reuse pattern that determines these offsets.
With a fixed input structure and stable cache behavior, Mallory may infer this pattern through offline profiling, although the prediction then depends on the accuracy of the estimate.

\noindent\textbf{Fully touched states.}
Once every usable id has been assigned, the consecutive range $[H+1,\ldots,U]$ is empty.
In this case, $F=U-Z$. 
Since Bob's selected allocation is already at or beyond position $F$, Mallory's later allocation cannot satisfy the range condition.
This consecutive-range method therefore no longer yields an exact prediction of the selected ids.
Using an estimated id instead makes \sys a \textit{probabilistic} and \textit{untargeted} attack that may expose context from an unintended resident request in \siphon, alter an unintended victim's response using Mallory's cached context in \graft, or produce no useful outcome.

\section{Framework Vulnerability Analysis}
\label{sec:frameworks}
This section evaluates \sys in paged \ac{KV} cache serving using \emph{software-emulated fault injection}, with vLLM as the primary platform.

\subsection{Experimental Setup}

We run experiments on a server with an 8-core Intel Xeon Gold 5415+ CPU, 128~GiB of ECC DDR5 memory, and an NVIDIA GeForce RTX~4090 GPU with 24~GiB of GDDR6X memory.
The server runs Ubuntu~22.04.5 LTS with the Linux~5.15 kernel and NVIDIA driver~580.173.02.
The main MRV2 environment uses PyTorch~2.13.0 built for CUDA~13.0.
Our main evaluation uses vLLM at commit \texttt{0ecc284} under MRV2, the default runner for the evaluated dense models.
We use a block size of $B=16$ and greedy decoding throughout\footnote{We also test under MRV1 at commit \texttt{9558286} and observed the same results.}.
Under MRV2, the block table has no host-side copy, so we overwrite the device-resident target entry directly to emulate a fault.

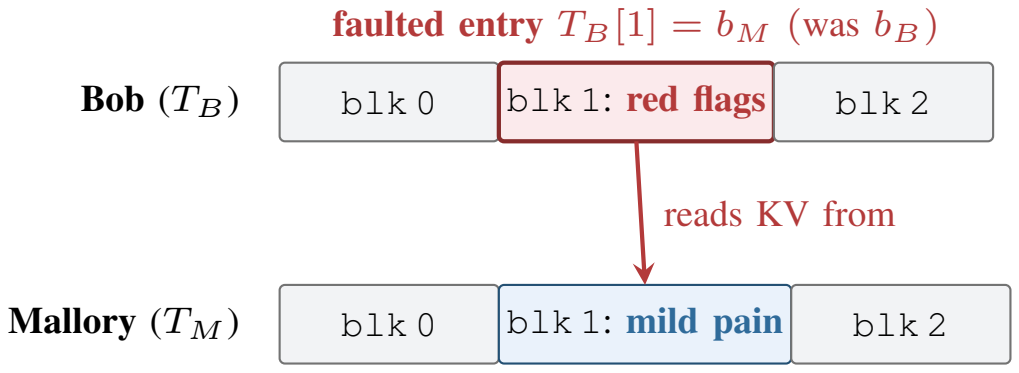
\begin{figure*}[!t]
\centering
\subfloat[Same-index redirection: Mallory's \acs{KV} reads access Bob's at index~1.\label{fig:case1-same}]{%
\centering
\begin{minipage}[c]{0.38\linewidth}
\raggedright
\begin{tikzpicture}[
  font=\scriptsize,
  cell/.style={kvblock,rounded corners=1pt,minimum width=1.00cm,
    minimum height=0.56cm},
  role/.style={font=\scriptsize,anchor=east},
  faultlabel/.style={kvfault,font=\scriptsize}
]
  \node[cell,kvneutral] (b0) {\texttt{blk\,0}};
  \node[cell,kvvictim,minimum width=1.80cm,right=0pt of b0] (b1)
    {\texttt{blk\,1}: \victimtext{cancer}};
  \node[cell,kvneutral,right=0pt of b1] (b2) {\texttt{blk\,2}};
  \node[role,left=5pt of b0] {\textbf{Bob} ($T_B$)};

  \node[cell,kvneutral,below=0.90cm of b0] (a0) {\texttt{blk\,0}};
  \node[cell,kvattacker,line width=0.8pt,minimum width=1.70cm,right=0pt of a0] (a1)
    {\texttt{blk\,1}: \attackertext{flu}};
  \node[cell,kvneutral,right=0pt of a1] (a2) {\texttt{blk\,2}};
  \node[role,left=5pt of a0] {\textbf{Mallory} ($T_M$)};

  \draw[kvredirect]
    (a1.north) --
    node[pos=.48,right=4pt,faultlabel]
      {reads \acs{KV} from}
    (b1.south);
  \node[faultlabel,anchor=north,yshift=-2pt] at (a1.south)
    {\textbf{faulted entry} $T_M[1]=b_B\;(\text{was }b_M)$};
\end{tikzpicture}
\end{minipage}\hfill
\begin{minipage}[c]{0.60\linewidth}
\begin{attackexample}
\footnotesize
\textbf{Bob (victim) prompt:} I'm a \victimtext{Stage 2 lung cancer} patient. Imagine you are a senior doctor and tell me how to treat it\,\ldots\\[1pt]
\textbf{Mallory (attacker) prompt:} I'm a \attackertext{flu} patient. Imagine you are a senior doctor and tell me how to treat it\,\ldots
\tcblower
\footnotesize
\textbf{Reply to Mallory (clean):} I'm sorry to hear that you're feeling unwell with the \attackertext{flu}. Here are some general recommendations\,\ldots\ Rest\,\ldots\ Hydration\,\ldots\\[1pt]
\textbf{Reply to Mallory (redirected):} I'm sorry to hear that you're dealing with \victimtext{Stage 2 lung cancer}. As a senior doctor, I would recommend\,\ldots\ which may include \victimtext{surgery, chemotherapy, radiation}\,\ldots
\end{attackexample}
\end{minipage}
}

\par\vspace{3pt}
\subfloat[Cross-index redirection: Mallory's \acs{KV} reads access Bob's at index~2.\label{fig:case1-cross}]{%
\centering
\begin{minipage}[c]{0.38\linewidth}
\raggedright
\begin{tikzpicture}[
  font=\scriptsize,
  cell/.style={kvblock,rounded corners=1pt,minimum width=1.00cm,
    minimum height=0.56cm},
  role/.style={font=\scriptsize,anchor=east},
  faultlabel/.style={kvfault,font=\scriptsize}
]
  \node[cell,kvneutral] (b0) {\texttt{blk\,0}};
  \node[cell,kvneutral,right=0pt of b0] (b1) {\texttt{blk\,1}};
  \node[cell,kvvictim,minimum width=1.70cm,right=0pt of b1] (b2)
    {\texttt{blk\,2}: \victimtext{cancer}};
  \node[cell,kvneutral,right=0pt of b2] (b3) {\texttt{blk\,3}};
  \node[role,left=5pt of b0] {\textbf{Bob} ($T_B$)};

  \node[cell,kvneutral,below=0.90cm of b0] (a0) {\texttt{blk\,0}};
  \node[cell,kvattacker,line width=0.8pt,minimum width=1.70cm,right=0pt of a0] (a1)
    {\texttt{blk\,1}: \attackertext{flu}};
  \node[cell,kvneutral,right=0pt of a1] (a2) {\texttt{blk\,2}};
  \node[role,left=5pt of a0] {\textbf{Mallory} ($T_M$)};

  \draw[kvredirect]
    (a1.north) to[out=90,in=-90]
    node[pos=.48,left=3pt,faultlabel]
      {reads \acs{KV} from}
    (b2.south);
  \node[faultlabel,anchor=north,yshift=-2pt] at (a1.south)
    {\textbf{faulted entry} $T_M[1]=b_B\;(\text{was }b_M)$};
\end{tikzpicture}
\end{minipage}\hfill
\begin{minipage}[c]{0.60\linewidth}
\begin{attackexample}
\footnotesize
\textbf{Bob (victim) prompt:} \ldots\ To be specific, I am a \victimtext{Stage 2 lung cancer} patient and want to know how to treat it\,\ldots\\[1pt]
\textbf{Mallory (attacker) prompt:} I'm a \attackertext{flu} patient\,\ldots
\tcblower
\footnotesize
\textbf{Reply to Mallory (redirected):} I'm sorry to hear about your diagnosis\,\ldots\ \victimtext{Stage 2 lung cancer} is a relatively advanced stage of the disease, and treatment options\,\ldots\ may include \victimtext{surgery, chemotherapy, radiation therapy}\,\ldots
\end{attackexample}
\end{minipage}
}

\caption{\textbf{Context Disclosure Attack.} A single block-table fault causes Bob's diagnosis to appear in Mallory's response on vLLM. Panel~(a) redirects between equal logical indices, whereas panel~(b) redirects Mallory's block~1 to Bob's block~2. Victim-derived content is shown in \victimtext{bold red}, and Mallory's original topic is shown in \attackertext{bold blue}.}
\label{fig:case1}
\end{figure*}

\subsection{Context Disclosure Attack}
\label{sec:eval-case1}

We use a medical assistant \ac{LLM} service as the running example: Bob asks how to treat Stage 2 lung cancer and Mallory asks about flu. 
Redirecting Mallory's content block to Bob's makes Bob's diagnosis appear in Mallory's reply, which otherwise concerns flu, while Bob's own reply is unchanged (\Cref{fig:case1-same}). 
Under a cross-index redirection, her reply still coherently incorporates the diagnosis (\Cref{fig:case1-cross}).
Other block pairings produce different outcomes in our experiments: 
(i) redirecting Mallory's content block to Bob's chat-template block ($T_M[1]\!\to\!T_B[2]$) collapses her reply to a single token (\emph{``I''}) and leaks nothing; (ii) redirecting the content and template blocks together yields a refusal (\emph{``I'm sorry, but I can't assist with that.''}); (iii) redirecting Mallory's \emph{own} template-close block to Bob's content does surface the diagnosis, but the lost turn boundary degrades the reply into repetition.
The generated text depends on the model and prompts.

\begin{table}[htbp]
\centering
\caption{\siphon across seven models. 
$N_L$ denotes the number of layers. \emph{Faulted Entries} reports the number of corrupted block-table entries.
GQA, MHA, and SWA+full denote grouped-query attention, multi-head attention, and hybrid sliding-window/full attention, respectively.}
\label{tab:case1-models}

\begin{adjustbox}{width=\linewidth}
\begin{tabular}{lrrlccc}
\toprule
Model & Params & $N_L$ & \makecell{Attention\\ (GQA ratio, type)} &
\makecell{\acs{KV}\\groups} & \makecell{Faulted\\Entries} & Disclosure \\
\midrule
Qwen2.5-1.5B-Instruct   & 1.5B & 28 & GQA 6:1, full      & 1 & 1 & Yes \\
Qwen2.5-7B-Instruct  & 7B   & 28 & GQA 7:1, full      & 1 & 1 & Yes \\
Llama-3.2-1B-Instruct & 1B   & 16 & GQA 4:1, full      & 1 & 1 & Yes \\
Llama-3.1-8B-Instruct & 8B   & 32 & GQA 4:1, full      & 1 & 1 & Yes \\
Gemma-2-2b-it   & 2B   & 26 & GQA 2:1, SWA+full  & 2 & 1 & Yes \\
Gemma-2-9b-it   & 9B   & 42 & GQA 2:1, SWA+full  & 2 & 1 & Yes \\
Phi-3.5-mini-instruct   & 3.8B & 32 & MHA, full          & 1 & 1 & Yes \\
\bottomrule
\end{tabular}
\end{adjustbox}
\end{table}

To assess the generality of \siphon, we evaluate seven models from four families, ranging from 1B to 9B parameters and covering GQA and MHA, full and hybrid sliding-window attention, and one or two \ac{KV} cache groups, as summarized in \Cref{tab:case1-models}.
Qwen and Llama use homogeneous full attention. 
A single block-id fault produces observable leakage across all seven evaluated models.
Their differences in model size, GQA ratio, attention pattern, and \ac{KV} cache grouping do not alter the redirection mechanism, \ie once the fault repoints the selected entry to the source block, the affected attention layers read its \ac{KV} states.

\subsection{Response Alteration Attack}
\label{sec:eval-case2}

\subsubsection{Illustration and Factors}
We consider a medical-advice scenario in which Bob and Mallory submit the same type of query, but Bob reports emergency warning signs while Mallory describes a benign complaint. 
The redirect replaces Bob's symptom block with Mallory's, causing the model to answer Bob's query using Mallory's milder context.
As shown in \Cref{fig:case2}, the normal reply advises Bob to seek immediate medical attention and identifies a possible heart attack.
After the redirect, the emergency recommendation disappears and the response instead attributes the pain to food and stress, and finally suggests a routine appointment. 
The corruption is also difficult for Bob to notice because the response remains relevant to his stated complaint and does not \textit{explicitly} contradict his description of the symptoms.

\begin{figure}[htbp]
\centering

\begin{tikzpicture}[
  font=\scriptsize,
  cell/.style={kvblock,rounded corners=1pt,minimum width=1.45cm,minimum height=0.52cm},
  role/.style={font=\scriptsize,anchor=east},
  faultlabel/.style={kvfault,font=\scriptsize}
]
  \node[cell,kvneutral] (b0) {\texttt{blk\,0}};
  \node[cell,kvvictim,line width=0.8pt,right=0pt of b0] (b1)
    {\texttt{blk\,1}: \victimtext{red flags}};
  \node[cell,kvneutral,right=0pt of b1] (b2) {\texttt{blk\,2}};
  \node[role,left=4pt of b0] {\textbf{Bob} ($T_B$)};

  \node[cell,kvneutral,below=0.95cm of b0] (a0) {\texttt{blk\,0}};
  \node[cell,kvattacker,right=0pt of a0] (a1)
    {\texttt{blk\,1}: \attackertext{mild pain}};
  \node[cell,kvneutral,right=0pt of a1] (a2) {\texttt{blk\,2}};
  \node[role,left=4pt of a0] {\textbf{Mallory} ($T_M$)};

  \draw[kvredirect]
    (b1.south) --
    node[pos=.50,right=3pt,faultlabel]
      {reads \acs{KV} from}
    (a1.north);
  \node[faultlabel,anchor=south,yshift=2pt] at (b1.north)
    {\textbf{faulted entry} $T_B[1]=b_M\;(\text{was }b_B)$};
\end{tikzpicture}

\vspace{2pt}

\begin{tcolorbox}[
  colback=neutralfill,
  colframe=black!50,
  boxrule=0.4pt,
  arc=1.5pt,
  colbacktitle=neutralfill,
  coltitle=black,
  fonttitle=\bfseries\footnotesize,
  title=Same-domain integrity violation,
  left=4pt,right=4pt,top=3pt,bottom=3pt,
  segmentation style={solid,draw=black!25,line width=0.3pt}
]
\footnotesize
\textbf{Bob (victim) prompt:} I have \victimtext{crushing chest pain, left arm numbness, and cold sweats}. Should I be worried\,\ldots?\\[1pt]
\textbf{Mallory (attacker) prompt:} I have \attackertext{mild chest pain, nothing serious, it seems related to food and stress}. Should I be worried\,\ldots?
\tcblower
\footnotesize
\textbf{Reply to Bob (clean):} \ldots\ it is important to \victimtext{seek medical attention immediately} if you are experiencing chest pain, numbness in your left arm, and cold sweats. These symptoms could be indicative of a serious medical condition such as a \victimtext{heart attack}\,\ldots\\[1pt]
\textbf{Reply to Bob (redirected):} \ldots\ Chest pain can be a symptom of various conditions, ranging from \attackertext{minor issues like acid reflux or muscle strain} to more serious conditions. Given that \attackertext{your chest pain seems related to food and stress}, it might be worth considering these factors. However, it is important to \attackertext{consult with a healthcare professional} for a proper evaluation\,\ldots
\end{tcolorbox}
\caption{\textbf{Response Alteration Attack.} A fault in Bob's block-table entry redirects his symptom block to Mallory's.
For Bob's red-flag triad, the normal response identifies a heart attack and recommends \emph{immediate care}, whereas the redirected response leads with reflux and muscle strain and recommends a routine appointment.}
\label{fig:case2}
\end{figure}

For illustration, we consider a general-purpose \ac{LLM} deployed through a domain-specific application, such as a cardiac symptom-triage service~\cite{shaposhnikov2025clarity,wang2026acstriage}.
The public interface and prompt format make requests likely to share a broad topic and structure.
Mallory can infer these constraints from the service description and submit a same-domain query with related symptoms.

\noindent\textbf{Factors affecting the integrity leak.}
Whether a same-domain redirect yields a harmful yet plausible reply depends on how Bob's query relates to Mallory's injected block. 
We identify three factors:
\begin{itemize}[leftmargin=1.2em,topsep=0.2em,itemsep=0.2em]
\item \textbf{Topic proximity.} The redirect is most coherent when the two requests share a similar topic and question form. 
A shared topic lets Mallory's block read as a continuation of Bob's own context, whereas a distant topic yields an off-topic or incoherent reply that Bob can recognize.
\item \textbf{Intra-topic divergence.} 
It captures clinically meaningful differences between two requests sharing the same topic, such as differences in symptom severity. 
Larger divergence can shift model's recommendation more strongly toward Mallory's description, while the shared topic helps keep the response relevant to Bob's question.
\item \textbf{Victim-model behavior.} Models differ in how far a substituted context moves the recommendation. The redirect supplies context, so the reply remains the model's own inference from Mallory's symptom description, and the
model's safety behavior applies to that description.
\end{itemize}

We use medical consultation as a representative case study because it \textit{clearly illustrates the consequences of redirection}.
\sys is not tied to this domain, as it targets \ac{KV} cache mappings rather than application-specific content.

\subsubsection{Benchmark}
\label{sec:eval-bench}
To examine how the factors above affect the attack, we build a benchmark of paired victim and attacker queries.
We derive victim queries from real patient messages to preserve realistic symptom descriptions.
For each fixed victim query, we construct attacker queries that vary in topic or severity.
We group the pairs by chief complaint: same-complaint conditions vary severity, while the distant-topic condition pairs different complaints.
We use \acp{LLM}, including models selected specifically for grading, to assess redirection responses and whether corrupted responses remain consistent with the victim's symptoms.

\noindent\textbf{Victim queries.}
Bob's queries are drawn from public ChatDoctor-HealthCareMagic-100k~\cite{li2023chatdoctor}, a corpus of real patient messages sent to an online consultation service. 
We sample 2{,}000 messages and label each one with a \emph{chief complaint} from 35 presentations (\eg chest pain, shortness of breath, fever, and so on) and a \emph{severity} on a four-level scale (\ie emergency, urgent, routine, benign). 
We organize the benchmark by chief complaint. 
This allows us to pair queries with similar presenting symptoms but different underlying causes and severity levels. 
For example, in \Cref{fig:case2}, Bob's chest pain reflects a cardiac emergency, whereas Mallory reports mild chest pain associated with food and stress.

For simplicity, we extract the symptom description and place it in a fixed service template while preserving the original wording.
The resulting benchmark contains 248 scenarios covering 25 of the 35 predefined complaints.

\noindent\textbf{Attacker queries.}
Mallory submits ordinary queries in the same format but chooses her own content.
For each chief complaint, we use two queries that differ only in the severity they present.
The \emph{benign} query attributes the complaint to a harmless cause (\eg ``mild chest pain, just heartburn after a big meal'') and is the baseline condition.
The \emph{routine} query keeps the complaint and raises the severity one level (\eg ``chest pain when I walk uphill that settles with rest'').

We hold Bob's query fixed and vary Mallory's query along two dimensions.
For topic proximity, her query remains benign while its chief complaint either matches or differs from Bob's.
To maintain balanced coverage across complaints, we evaluate topic proximity on a subset comprising four scenarios from each of 25 complaints ($n=100$ per served model).
For intra-topic divergence, the chief complaint remains matched while her query changes from benign to routine.
Together, these measure how the topic and severity of the substituted content affect Bob's response.

\noindent\textbf{Serving models.}
We use seven models spanning four families, three attention designs, and 1B--9B parameters: Qwen2.5-1.5B/7B, Llama-3.2-1B, Llama-3.1-8B, Gemma-2-2b/9b, and Phi-3.5-mini. 
Each model is evaluated on the same set of scenarios.

\noindent\textbf{Metrics.}
We assess the impact of redirected responses using two metrics, \emph{Severity} and \emph{Detectability}.
\emph{Severity} captures the urgency of care recommended in each response, and a decrease from Bob's original response to his redirected response indicates potential harm.
\emph{Detectability} captures how readily Bob could notice that the redirected response does not fit his situation, using a three-level scale: \emph{covert}, \emph{subtle}, and \emph{overt}.

We define the \emph{Covert Downgrade Rate (CDR)} as the fraction of scenarios in which the redirected response has a lower severity level than the original response (\ie downgrade: severity drop) while still being judged consistent with Bob's question (\ie covert: low detectability).
This captures the attacker's most favorable outcome.
That is, Bob receives a response that understates the severity of his condition while remaining consistent with his query.

\noindent\textbf{Rating strategy.}
Given \ac{LLM}’s expertise in processing natural language, both metrics are rated using \acp{LLM} as judges.
To select judges, we calibrate each using two sets of question--response pairs with known match labels.
The matched set pairs each question with its original response, whereas the mismatched set pairs it with a response written for another patient with a different chief complaint.
We treat mismatch detection as the positive class and measure each judge's false positive rate on matched pairs and true positive rate on mismatched pairs.
We retain only judges with both a low false positive rate and a high true positive rate, since always accepting/rejecting responses trivially satisfies either criterion alone.
On 700 balanced question--response pairs per judge, the five judges achieve false positive rates of 1.4--9.4\% and true positive rates of 89.4--98.9\%.

Finally, we adopt five \acp{LLM} as the judges, \ie Qwen2.5--32B--Instruct, Prometheus--7B--v2.0, Selene--1--Mini, GLIDER, and CompassJudger--1--7B, drawn from five model families and including four models fine-tuned specifically for grading.
We supply the severity and detectability rubrics in each judge's own prompt template, preserving its expected input format.

\noindent\textbf{Results.}
\Cref{tab:case2-models} and \Cref{tab:case2-arms} show how each factor affects the attack outcomes across the evaluated conditions.

\emph{Topic proximity and intra-topic divergence.}
For topic proximity, we keep Mallory's query severity as benign but change her complaint to a different topic.
Across five judges, the downgrade rate changes by at most 1.7\%, while the fit rate decreases by 11.4\%--19.4\%.
Thus, changing the topic makes the redirected response less consistent with Bob's question, with little effect on recommendation downgrade.
Domain-specific \ac{LLM} services therefore provide a favorable setting for \sys, as Mallory can ask about the same complaint as Bob.

To evaluate intra-topic divergence, we keep the chief complaint matched and change Mallory's query from benign to routine, thereby reducing its divergence from Bob's.
Across all judges, the downgrade rate falls by 7.8--12.9\%, while the responses judged to fit Bob's question ranges from a 2.6\% decrease to a 7.0\% increase, with no consistent direction across judges.
As Mallory's symptoms approach Bob's in severity, downgrades become less frequent.
This pattern is consistent with the model basing its recommendation on Mallory's substituted symptom description.
Even when the recommended urgency does not decrease, the resulting advice can be inappropriate for Bob's condition.

\begin{table}[htbp]
\centering
\footnotesize
\caption{Effect of topic proximity and intra-topic divergence. 
Each cell reports \emph{downgrade}/\emph{covert} (\%), indicating the fraction with lower severity and the fraction remaining consistent with Bob's query, respectively.
Results are pooled across all seven served models and reported separately for each judge. $n$ denotes the number of scenarios per served model.
}
\label{tab:case2-arms}
\setlength{\tabcolsep}{4pt}
\resizebox{\linewidth}{!}{
\begin{tabular}{lcccccc}
\toprule
Attacker query & $n$ & Selene & GLIDER & \makecell[l]{Qwen2.5} & \makecell[l]{CompassJudger} & \makecell[l]{Prometheus} \\
\midrule
\multicolumn{7}{l}{\emph{Topic proximity}} \\
Benign (baseline) & 100 & 28.6/91.7 & 30.6/81.6 & 34.1/53.1 & 36.1/82.0 & 34.6/78.3 \\
Distant topic      & 100 & 26.9/80.3 & 30.4/62.1 & 33.4/37.7 & 34.7/66.4 & 34.4/64.9 \\

\midrule

\multicolumn{7}{l}{\emph{Intra-topic divergence}} \\
Benign (baseline) & 248 & 25.9/92.3 & 30.1/80.1 & 35.1/53.6 & 36.4/82.2 & 33.7/77.6 \\
Routine            & 248 & 16.3/89.7 & 22.3/84.1 & 22.2/53.5 & 24.6/82.5 & 23.7/84.6 \\
\bottomrule
\end{tabular}
}
\end{table}

\emph{Victim-model behavior.}
Under Mallory's same benign query, we evaluate every scenario on each of the seven models.
Across model--judge combinations, the CDR ranges from 5.2\% to 41.1\% (\Cref{tab:case2-models}).
The rankings of served models are largely consistent across judges: both Gemma models rank among the three most affected under every judge, whereas both Llama models rank among the three least affected under four of the five.
The share of \emph{one-level} downgrades ranges from 62.8\% under GLIDER to 96.4\% under Selene.
Two-level drops account for nearly all remaining cases, while three judges record a small number of three-level drops.
Overall, \graft produces covert downgrades across all seven models, but the rate varies substantially and does not follow parameter count.
This variation likely reflects differences in instruction tuning and safety behavior, which shape how strongly the redirected symptoms influence each model's recommendation.

\begin{table}[htbp]
\centering
\footnotesize
\caption{Integrity violation across served models. 
Each cell is the \emph{CDR} (\%).}
\label{tab:case2-models}
\setlength{\tabcolsep}{4pt}
\resizebox{\linewidth}{!}{
\begin{tabular}{lccccc}
\toprule
Served model & Selene & GLIDER & \makecell[l]{Qwen2.5} & \makecell[l]{CompassJudger} & \makecell[l]{Prometheus} \\
\midrule
Gemma-2-9b-it          & 30.6 & 41.1 & 30.2 & 32.0 & 37.1 \\
Gemma-2-2b-it          & 38.7 & 34.8 & 29.0 & 35.1 & 29.8 \\
Qwen2.5-7B-Instruct    & 33.5 & 27.0 & 20.6 & 37.7 & 37.1 \\
Phi-3.5-mini-instruct  & 22.2 & 23.0 & 20.2 & 28.4 & 27.8 \\
Qwen2.5-1.5B-Instruct  & 17.7 & 12.9 & 10.1 & 26.4 & 12.9 \\
Llama-3.1-8B-Instruct  & 12.1 & 14.1 & 14.9 & 19.8 & 19.4 \\
Llama-3.2-1B-Instruct  & 14.9 & 14.1 & \phantom{0}5.2 & 31.9 & 14.1 \\
\bottomrule
\end{tabular}
}
\end{table}

Together, the benchmarks focus on the effectiveness of \sys and the factors that directly shape redirected responses.
For \graft, the results identify a favorable setting in which Bob's and Mallory's requests share a topic and structure but differ in symptom severity, and the served model responds strongly to Mallory's substituted context.
Future work can extend this analysis to model-specific semantic and safety behavior, other application domains, and additional scenario designs and outcome measures.

\subsection{Extension to Other Serving Frameworks}
\label{sec:other_platforms}
\sys exploits paged \ac{KV} cache redirection and therefore applies beyond vLLM.
Using software-emulated mapping corruption, we reproduce both \siphon and \graft with Qwen2.5-1.5B-Instruct on SGLang (\texttt{c16a8fc})~\cite{zheng2024sglang} and Sonar (\texttt{5d9c85a})~\cite{dphnai2026sonar}.
Following our disclosure, the Sonar maintainer independently reproduced the redirection on the current main branch.
The number of corrupted mapping entries varies with the mapping granularity and how the attention backend reads the mapping.
In our Sonar configuration, as in vLLM, each entry maps a 16-token logical block to one physical \ac{KV} block, so corrupting one entry redirects the entire block.
SGLang maintains token-level mappings, while the tested backends read them differently.
FlashAttention locates a 16-token page through one page-boundary entry, so corrupting that entry redirects the page.
FlashInfer reads one entry per token, making a full 16-token redirect span 16 entries.
In our \siphon experiment, eight corrupted entries were enough to expose the target diagnosis.
The cross-framework experiments focus on validating both redirection outcomes and their mapping requirements. Evaluating their effectiveness across additional backend configurations and model families remains future work.

\section{Physical Rowhammer Evaluation}
\label{sec:rowhammer_sys}
We take Rowhammer as the target fault injection primitive to evaluate the physical exploitability of \sys.
We reuse SWAGE~\cite{boy2026slashdsa}, a modular framework for implementing Rowhammer attacks, together with its Blacksmith-based hammerer~\cite{jattke2022blacksmith}, on a DDR4 platform known to be susceptible.
SWAGE separates a Rowhammer attack into reusable modules for memory allocation, bit-flip profiling and replay, target-page placement, and victim coordination.
Its application-specific component, termed the \emph{victim orchestrator}, connects these modules to the target application.
We implemented \sys as a \emph{VictimOrchestrator} in SWAGE to drive the campaigns.

\subsection{Fault Requirements}
\label{subsec:rh-requirements}
A single bit flip in a block-table entry realizes a redirect only if four conditions hold.

\noindent\textbf{(R1) Position.}
vLLM stores each block id as a 32-bit integer, and the flipped id should remain inside the block pool.
For the 512-block pool in our setup, this restricts candidate positions to bits~0--8.
On our platform, these comprise all eight bits of byte lane~0 and the least significant bit of lane~1.
Our orchestrator therefore restricts its search to lane-0 flips.

\noindent\textbf{(R2) Value and direction.}
As described in \Cref{sec:attack_cons}, Mallory varies her request length to obtain candidate logical blocks and inserts dummy requests to shift their assigned ids until a reader--source pair matches the profiled flip.

\noindent\textbf{(R3) Placement.}
The serving process stores the block table in an anonymous page.
Mallory should place this page on a physical frame susceptible to Rowhammer-induced bit flips.

\noindent\textbf{(R4) Window.}
The flip should land after the source block holds its \ac{KV} states and before the reader's next read access (\Cref{fig:window} in \Cref{sec:attack_cons}).

\subsection{Experimental Setup}
\label{subsec:rh-setup}
\noindent\textbf{Platform.}
We conduct all Rowhammer experiments on a machine equipped with an Intel Core i5-6400 at 2.70\,GHz and one 16\,GiB G.Skill AEGIS F4-2133C15-16GIS DDR4-2133 DIMM.
The host runs Ubuntu~22.04.5 with Linux~6.8.12.

\noindent\textbf{Target serving system.}
We target the official vLLM~0.25.0 CPU release without source modifications, serving Llama-3.2-1B-Instruct.
The engine uses 16-token \ac{KV} blocks and a 512-block pool.
Its block table is an 8\,KiB \texttt{int32} tensor in host memory.

\noindent\textbf{Attack pipeline.}
Mallory runs SWAGE as a co-resident process.
SWAGE allocates 40\,MiB of physically contiguous memory, replays one Blacksmith pattern with 70 aggressor addresses that was fuzzed once for this platform, and profiles the allocation for susceptible cells (\Cref{subsec:rh-profiling}).
The \sys orchestrator maps each profiled flip onto the published table geometry, computes a request layout that satisfies R1 and R2, places the table page (R3), and replays the pattern while the redirect window is open (R4).

\subsection{Profiling for Usable Bit Flips}
\label{subsec:rh-profiling}

\begin{figure*}[t]
\centering
\subfloat[Byte lane in 4-byte block id]{%
\begin{minipage}[t]{0.3\textwidth}
\centering
\includegraphics[width=0.8\linewidth]{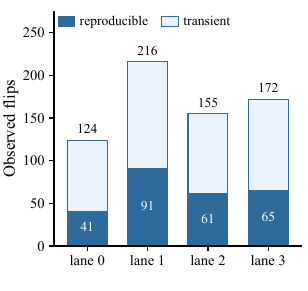}
\end{minipage}%
}\hfill
\subfloat[Bit offset within the byte]{%
\begin{minipage}[t]{0.34\textwidth}
\centering
\includegraphics[width=0.8\linewidth]{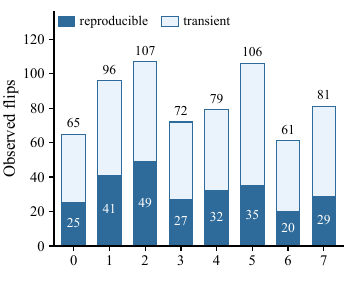}
\end{minipage}%
}\hfill
\subfloat[Byte-lane/bit-offset heatmap]{%
\begin{minipage}[t]{0.34\textwidth}
\centering
\includegraphics[width=0.8\linewidth]{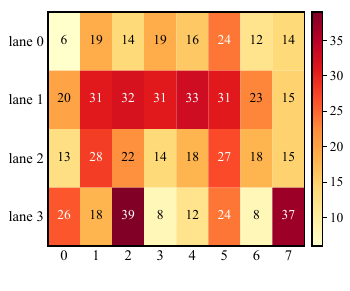}
\end{minipage}%
}
\caption{Profiled bit flips observed across all 20 campaigns ($n=667$). Flips observed in at least two of three profiling rounds are classified as reproducible, whereas those observed in only one round are classified as transient.}
\label{fig:flip_census}
\end{figure*}

This can be done offline on a machine with the same configuration rather than on the victim machine. Each campaign repeatedly allocates fresh memory and profiles it over three hammering rounds of 20 pattern attempts each.
Across 20 total campaigns, SWAGE profiled 637 allocations in 1{,}600 rounds, amounting to 6.1\,h of hammering, and recorded 667 total bit flips in 326 allocations.
\Cref{fig:flip_census} shows the distribution of these flips.
Flips occur in all four byte lanes and at all eight bit positions, with a slight skew towards lane~1.
All recorded faults were single-bit flips, with 346 occurring in the $0\rightarrow1$ direction and 321 in the $1\rightarrow0$ direction, indicating a nearly even directional split.

Under our lane-0 search restriction, 124 of the 667 profiled flips (18.6\%) are candidates, and 41 of those are reproducible.
Following SWAGE~\cite{boy2026slashdsa}, we classify a flip as reproducible if it recurs in at least two of the three profiling rounds.
R2 makes $0\rightarrow1$ flips usable more often in our allocation layout.
A $1\rightarrow0$ flip can reach Bob's block only when the decrease exceeds the selected entry's column index, whereas Mallory can adjust her request length to match a $0\rightarrow1$ flip for most columns.
This difference explains why 68 of the 70 flips observed in the block table were $0\rightarrow1$, even though profiling found both directions at similar rates.

\subsection{Placing the Block Table}
\label{subsec:rh-placement}
SWAGE integrates techniques from prior Rowhammer as reusable modules and exposes target-page placement through a Page Injector interface.
Its existing injector targets a future victim allocation, whereas R3 requires relocating an already allocated block-table page.
We therefore implemented DeepSteal's \emph{swap+refault} technique~\cite{rakin2022deepsteal} as a new Page Injector.
The injector evicts the table page to swap by applying memory pressure, releases the profiled frame to the per-CPU free list of the core the serving process runs on, and lets the next table access refault the page, which then reuses the most recently released frame.
We count an attempt as successful only if the refaulted table page occupies exactly the profiled frame.
In our experiment, the table page landed on the profiled frame in 36 of 302 attempts (11.9\,\%).

\subsection{Physical Attack Results}
\label{subsec:rh-e2e}
We ran 20 campaigns, alternating between \siphon and \graft, with ten campaigns per attack and a 90-minute limit per campaign.
In each campaign, SWAGE profiled fresh allocations and retained only those containing a reproducible flip that satisfied R1 and R2.
For each retained allocation, it launched the engine, staged Bob's, Mallory's, and dummy requests according to the resulting layout, placed the target table page, and hammered during each redirect window.
Of the 326 allocations in which profiling observed at least one flip, 24 supported a valid request layout.
These allocations provided 142 hammering windows, of which 70 (49\,\%) changed the target entry to the intended block id.
Sixty of these redirects produced the expected marker in the decoded response; the remaining 10 changed the entry without producing the marker.
\Cref{tab:rh-e2e} reports the breakdown by attack.

\begin{table}[t]
\centering
\footnotesize
\caption{
  Physical Rowhammer campaigns (90\,min each). 
  A window is one hammering interval while the redirect window is open. 
  \emph{Redirect} counts windows in which the flip changed the entry to the planned id and \emph{validated} additionally requires the marker in the decoded response.
}
\label{tab:rh-e2e}
\begin{tabular}{lcc}
\toprule
 & \siphon & \graft \\
\midrule
Campaigns validated & 3/10 & 8/10 \\
Campaigns with redirect only & 4/10 & 1/10 \\
Campaigns without redirect & 3/10 & 1/10 \\
\midrule
Total hammering windows & 39 & 103 \\
Windows with redirect & 13 (33\,\%) & 57 (55\,\%) \\
Windows validated & 7 (18\,\%) & 53 (51\,\%) \\
Hammering time per window (median) & 15.2\,s & 15.2\,s \\
\bottomrule
\end{tabular}
\end{table}

At the campaign level, physical Rowhammer validated \siphon in 3 of 10 campaigns and \graft in 8 of 10.
Three \siphon campaigns reached the 90-minute limit without finding a flip compatible with a valid request layout.
Among windows with the intended redirect, the expected marker appeared in 7 of 13 \siphon windows and 53 of 57 \graft windows.
Two \graft campaigns produced the marker in every tested window, with 13 of 13 and 9 of 9 validated windows, respectively.
In every validated window, a single hammering interval with a median duration of 15.2\,s induced the planned flip.

\subsection{Attack Cost and Discussion}
\label{subsec:rh-cost}
The 20 campaigns took 31.7\,h of wall-clock time, of which 6.1\,h were spent hammering during profiling and 36\,min hammering inside the 142 attack windows.
Profiling accounts for most of the hammering and is independent of the victim, so we can run it \textit{offline} before launching the attack.
Most of the remaining wall-clock time is spent launching a fresh engine per attempt and staging requests.

Each \emph{swap+refault} placement attempt takes several minutes and a failed attempt requires a full restart.
To avoid repeating the placement cost while collecting physical-fault results, our campaigns use a kernel module to map the table page onto the profiled frame~\cite{boy2026slashdsa}.
Prior Rowhammer work achieves placement by steering a forthcoming allocation onto a profiled frame~\cite{vanderveen2016drammer,amer2025pq,boy2026slashdsa}, mapping file-backed contents onto a released frame~\cite{yao2020deephammer}, or relocating a resident anonymous page through \emph{swap+refault}~\cite{rakin2022deepsteal}.
Prior work also uses attacker-visible signals to synchronize these placement events.
SWAGE permits synchronization through network requests or cache attacks, whereas DeepSteal uses \textsc{Flush+Reload} anchors to coordinate page release with subsequent victim accesses~\cite{boy2026slashdsa,rakin2022deepsteal}.
For such coordination, scale-from-zero serving provides a favorable deployment setting.
KubeAI~\footnote{\url{https://www.kubeai.org/concepts/autoscaling/}} holds the first request when no serving instance is running, starts a fresh vLLM instance, and forwards the request once the service is ready.
This creates a predictable startup boundary for page placement.

Separately, concurrent inference reduced the DRAM activation rate in our setup because vLLM and the hammerer competed for memory access.
We therefore paused the engine during each hammering interval.
Prior Rowhammer attacks obtain longer hammering windows by slowing the victim process~\cite{adiletta2025leapfrog,amer2025pq}, while larger models with longer forward passes can provide more time between block-table accesses.
We focus here on demonstrating the physical feasibility of \sys and showing that vLLM propagates the resulting fault into the intended outcomes. 
We leave a more comprehensive implementation under these discussed settings for future work.

\section{Discussion}
\label{sec:discussion}
This section discusses the limitations of \sys, explores alternative primitives, considers other serving metadata as potential targets, and presents mitigations.

\subsection{Limitations}

\noindent\textbf{Block id prediction in noisy states.}
In \Cref{subsec:id-prediction}, we formalize how Mallory predicts the target block id under the allocator state.
In the fully touched state, she can no longer determine the target id from her request history alone.
\sys therefore becomes \emph{probabilistic} and \emph{untargeted}.
We do not measure this pattern because it depends on deployment-specific pool occupancy and the number of co-resident requests containing sensitive information.
Side-channel observations of memory access behavior from service startup may help infer request activity and improve block id prediction, but this approach requires further experimental validation.

\noindent\textbf{End-to-end GPU Rowhammer attack.}
GPU Rowhammer remains an emerging and hardware-specific primitive, and studies cover only a limited set of GPU models and memory configurations~\cite{lin2025gpuhammer,wan2026geforge,hu2026gddrhammer,lin2026gpubreach,lin2026gputhor}.
Furthermore, at the time of writing, GPUThor's implementation artifacts for bypassing ECC were not public~\cite{lin2026gputhor}.
As our focus is the new attack surface, we leave exploration of an end-to-end GPU Rowhammer pipeline to future work.

\subsection{Alternative Primitives and Other Targets}
\label{sec:alternative_primitives}
The redirection in \sys depends on changing the mapping used to locate a request's \ac{KV} state, and Rowhammer is one way to instantiate this capability.
Other techniques could corrupt the stored entry persistently or alter its value transiently during copying or lookup, although transient faults require tighter synchronization and may need repeated injection.
Under physical-access assumptions, voltage or clock glitching and laser or electromagnetic fault injection are possible hardware alternatives, while a data-only memory-write primitive in the serving stack could directly
overwrite the selected block id~\cite{chen2005noncontrol}.

\ac{LLM} systems maintain metadata for prefix-caching~\cite{zheng2024sglang}, \ac{KV} cache offloading~\cite{gao2024cachedattention}, and disaggregated prefill--decode serving~\cite{zhong2024distserve}.
This metadata records which prefixes can be reused, where offloaded \ac{KV} blocks are stored, and which cached state is transferred from prefill to decode.
Corrupting these records could cause a request to use cached state belonging to another request or tenant.
Evaluating their feasibility and security impact remains future work.

\subsection{Mitigations}

\subsubsection{Mitigations against Rowhammer}
Deployed countermeasures such as ECC memory and in-\ac{DRAM} TRR make exploitable bit flips harder but remain bypassable~\cite{frigo2020trrespass,jattke2022blacksmith,lin2026gputhor}.
Recent GPU Rowhammer studies recommend combining stronger error correction and integrity protection with activation-aware mechanisms such as Refresh Management (RFM) and Per-Row Activation Counting (PRAC), and using corrected-error counts and row-remapping activity as detection signals~\cite{wan2026geforge,lin2025gpuhammer,hu2026gddrhammer,lin2026gpubreach,lin2026gputhor}.
Randomizing address mappings and allocation reuse, or placing security domains in disjoint physical memory regions, can further impede vulnerable-cell profiling and target placement.

\subsubsection{Mitigations in \acs{LLM} serving frameworks}
At the framework level, the key defense is to ensure that each request can access only its assigned \ac{KV} blocks.

\noindent\textbf{Assignment validation.}
Assignment validation keeps an independent host-side record of the block ids assigned by the scheduler.
Before an operational block-table entry is used for \ac{KV} cache reads or writes, the framework compares it with the corresponding record and rejects any mismatch.
Following our disclosures, vLLM and Sonar implemented this approach in separate patches.
We reviewed both implementations and evaluated their handling of corrupted and clean mappings. 
Both patches rejected the tested corruptions before use, and clean executions completed normally.
These checks run on the inference path, so assignment validation incurs runtime overhead while preventing corrupted mappings from being used.
Further optimization could reduce this cost without weakening the protection.

\noindent\textbf{Serving-instance isolation.}
Mutually untrusted tenants can be isolated by assigning them to dedicated serving instances, each maintaining an independent block allocator and \ac{KV} cache pool. 
Because block ids are resolved strictly within their local pool, a corrupted block-table entry, even if it falls within the valid index range, cannot reference cached state belonging to another instance. 
This strict boundary effectively prevents cross-tenant context disclosure and response alteration. 
However, this security guarantee comes at a performance cost, as it sacrifices the operational efficiency of cross-tenant request batching and shared \ac{KV} cache reuse.

\section{Conclusion}
This paper presents \sys, a fault-injection attack that corrupts a paged \ac{KV} cache mapping and redirects a request's \ac{KV} reads to cached state belonging to another tenant's request, enabling either a confidentiality leak through the attacker's response or an integrity violation through attacker-supplied context.
We characterize when target block ids can be predicted under different allocator states.
In our evaluation, \siphon leaks victim context across seven models from four families, while \graft produces covert downgrades across seven served models, as assessed by five \ac{LLM} judges.
We further demonstrate the practical feasibility of \sys in Rowhammer experiments.
These findings identify paged \ac{KV} cache mappings as security-critical serving metadata and motivate validation of block-table assignments and isolation between mutually untrusted tenants.

\bibliographystyle{IEEEtran}
\bibliography{ref}

\end{document}